\pdfoutput=1

\documentclass[12pt]{iopart}

\usepackage{iopams}

\usepackage{revsymb4-1} 
\usepackage{graphicx}
\usepackage{lscape}  

\newcommand{\vect}[1]{\vec{#1}}

\begin{document}

\title{Simulations of atomic trajectories near a dielectric surface}

\author{N. P. Stern$^1$, D. J. Alton$^1$, and H. J. Kimble$^1$}

\address{$^1$ Norman Bridge Laboratory of Physics 12-33,
California Institute of Technology, Pasadena, California 91125, USA}
\ead{hjkimble@caltech.edu}

\begin{abstract}
We present a semiclassical model of an atom moving in the evanescent field of a microtoroidal resonator.  Atoms falling through whispering-gallery modes can achieve strong, coherent coupling with the cavity at distances of approximately 100 nanometers from the surface; in this regime, surface-induced Casmir-Polder level shifts become significant for atomic motion and detection.  Atomic transit events detected in recent experiments are analyzed with our simulation, which is extended to consider atom trapping in the evanescent field of a microtoroid.

\end{abstract}

\pacs{ 37.30.+i , 34.35.+a, 42.50.Ct, 37.10.Vz}

\maketitle

\section{Introduction}

Strong, coherent interactions between atoms and light are an attractive resource for storing, manipulating, and retrieving quantum information in a quantum network with atoms serving as nodes for quantum processing and storage and with photons acting as a long-distance carrier for communication of quantum information~\cite{Kimble:2008}.  One realization of a quantum node is an optical cavity, where light-matter interactions are enhanced by confining optical fields to small mode volumes.  In the canonical implementation, a Fabry-Perot resonator with intracavity trapped atoms enables a panoply of cavity quantum electrodynamics (cQED) phenomena using single photons and single atoms, and thereby, validates many aspects of a cQED quantum node~\cite{Miller:2005, Wilk:2007}.

Despite these achievements, high-quality Fabry-Perot mirror cavities typically require significant care to construct and complex experimental instrumentation to stabilize.  These practical issues have begun to be addressed by atom chips~\cite{Reichel:2002, Folman:2002}, in which atoms are manipulated in integrated on-chip microcavity structures offering a scalable interface between light and matter~\cite{Vahala:2003, Colombe:2007, Gehr:2010}.  Owing to their high quality factors, low mode volumes, and efficient coupling to tapered optical fibers~\cite{Armani:2003}, microtoroidal resonators are a promising example of microcavities well-suited for on-chip cQED with single atoms and single photons~\cite{Spillane:2005}.  Strong coupling~\cite{Aoki:2006, Alton:2010} and non-classical regulation of optical fields~\cite{Dayan:2008, Aoki:2009} have been demonstrated with atoms and the whispering-gallery modes of a silica microtoroidal resonator.

In our experiments with microtoroids, Cs atoms are released from an optical trap and fall near a silica toroid, undergoing coherent interactions with cavity modes as each atom individually transits through the evanescent field of the resonator.  In the most recent work of~\cite{Alton:2010}, atom transits are triggered in real-time to enable measurement of the Rabi-split spectrum of a strongly-coupled cQED system.  Whereas a single atom is sufficient to modify the cavity dynamics, falling atoms are coupled to the cavity for only a few microseconds.  Atom dropping experiments necessarily involve a large ensemble of individual atomic trajectories and represent, consequently, a far more complex measurement result.

Interactions between a neutral atom and a dielectric surface modify the radiative environment of the atom resulting in an enhanced decay rate~\cite{Lukosz:1977} and Casimir-Polder (CP) forces~\cite{Sukenik:1993, Bordag:2001}.  These \emph{perturbative} radiative surface interactions are usually insignificant in cQED experiments with Fabry-Perot resonators where atoms are far from mirror surfaces, but in microcavity cQED, atoms are localized in evanescent fields with scale lengths $\lambda/2\pi \sim 150$ nm near a dielectric surface.  The experimental conditions for microtoroidal cQED with falling atoms in~\cite{Alton:2010} necessarily involve significant CP forces and level shifts while simultaneously addressing strong coupling to optical cavity modes.  Theoretical analysis of this experiment requires addressing both the strong atom-cavity interactions and atom interactions with the dielectric surface of the microtoroid. As reported in~\cite{Alton:2010}, spectral and temporal measurements offer signatures of both strong coupling to the cavity mode and the significant influence of surface interactions on atomic motion.  The role of these effects is quantified with detailed simulation of the trajectories of falling atoms detected in the real-time at low photon numbers.

In this article, we discuss in detail the approach used to simulate atomic motion near the surface of an axisymmetric dielectric resonator under the influence of strong coherent interactions with cavity modes.  The experimental detection method of~\cite{Alton:2010} is implemented stochastically in a semiclassical simulation of atom trajectories.  These simulations provide a perspective on the atomic motion of atom transits recorded in our microtoroid experiments, while offering additional insights into the loading of optical evanescent field traps.  In section~\ref{Sec:theory}, we outline the semiclassical model of a two-level atom coupled to the whispering gallery modes of a microtoroidal resonator.  In section~\ref{sec:OpticalForce}, we review the optical dipole forces which are a critical factor influencing atomic motion in an optical cavity.  Our calculations of modified emission rates and Casimir-Polder surface interactions are detailed in section~\ref{sec:Surface}.   Section~\ref{sec:exp} describes the implementation of our model for simulating recent atom-toroid experiments.  Finally, section~\ref{sec:trap} extends our simulation to evanescent field traps around a microtoroid.

\section{Atoms in a microtoroidal cavity}\label{Sec:theory}

We approach the motion of atoms moving under the influence of surface interactions and coherent cavity dynamics with a semiclassical method to efficiently simulate a large number of atom trajectories.  For surface interactions, dispersion forces are calculated \emph{perturbatively} using the linear response functions of SiO$_2$ and a multi-level atom.  For nearly-resonant \emph{non-perturbative} coherent interactions between atom and cavity, the atomic internal state and the cavity field are treated quantum mechanically within the two-level and rotating-wave approximations.

Simulations of atomic motion follow the semiclassical method detailed in~\cite{Doherty:1997}. Mechanical effects of light are incorporated classically as a force $\vect{F}(\vect{r})$ on a point particle atom at location $\vect{r}$.   Trajectories $\vect{r}(t)$ are calculated with a Langevin equation approach to incorporate momentum diffusion from fluctuations.  At each simulation time step $t^i$, the atomic velocity is calculated as:
\begin{equation}\label{eq:langevin}
v^{i+1}_j = v^{i}_j + F^i_j \Delta t/m_{\rm Cs} + \sqrt{2 D^i_{jj} \Delta t/m_{\rm Cs}^2} W^i_j
\end{equation}
where $\vect{v}^i$ is the velocity at the $i$ time step, $m_{\rm Cs}$ is the atomic mass, and $\Delta t$ is the simulation time step $t^{i+1}-t^{i}$. The $\vect{W}^i$ are normally distributed with zero mean and standard deviation of 1.  Given the force $\vect{F}$ and diffusion tensor $D_{ij}$ as discussed in section~\ref{sec:OpticalForce}, the atom trajectory $\vect{r}(t)$ and cavity transmission and reflection coefficients, $T(t)$ and $R(t)$ are calculated.  A single atom strongly coupled to the cavity mode has a large effect on cavity fields and optical forces, requiring simultaneous solutions of atomic motion and cQED dynamics.

Full quantization of atomic motion leads to an unwieldy Hilbert space not conducive to efficient simulation. In contrast, semiclassical methods are well-suited for simulating atomic motion in experiments with falling atoms near resonators.  The ratio of the recoil energy to the linewidth of the cesium $6S_{1/2}\rightarrow 6P_{3/2}$ transition is less than $10^{-3}$. Further, the recoil velocity of $\sim 3.5$ mm/s is much less than the typical velocity of falling atoms of order $200$ mm/s so that each spontaneous emission event represents a small momentum kick.  Cavity fields and internal atomic states respond quickly to environment changes, allowing calculation of optical forces and momentum diffusion in a constant-velocity limit at time $t$ and energy shifts from surface interactions as if atom the atom were stationary.  The remainder of this section discusses the quantum mechanical equations of motion for the atom and cavity fields in the low-probe intensity limit, to be followed later by contributions to the force $\vect{F}$ used in~\eref{eq:langevin}.

\subsection{Modes of a microtoroidal resonator}\label{subsec:modes}

\begin{figure}[tpb]
  \begin{center}
    \includegraphics{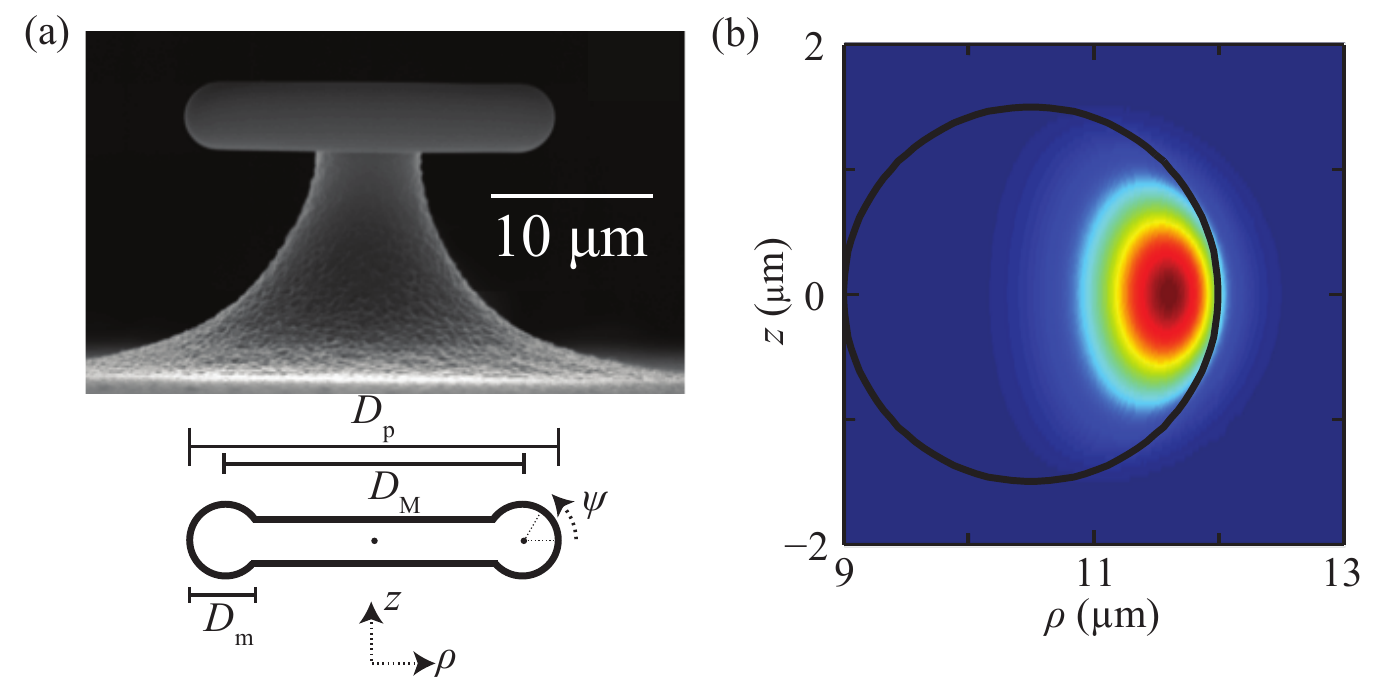}
  \end{center}\vspace{-0.5cm}
\caption{\label{Fig-modes} (a) A scanning electron microscope image of a microtoroid with definitions of the relevant parameters discussed in the text. (b) The lowest order mode function $f(\rho, z)$ of a toroid mode with $\{D_{\rm p}, D_{\rm m}\} = \{24, 3\}$ $\mu$m and $m = 118$ and $\lambda = 852$ mn.  }
\end{figure}

An idealized microtoroid has axial symmetry, so we work in a standard cylindrical coordinate system $\vect{r}\rightarrow (\rho, \phi, z)$.  The toroid is modeled as a circle of diameter $D_{\rm m}$ with dielectric constant $\epsilon$ revolved around the $z$-axis to make a torus of major diameter $D_{\rm M}$ (Fig.~\ref{Fig-modes}(a)).  The toroid is therefore defined by its minor diameter $D_{\rm m}$ and its principal diameter $D_{\rm p} = D_{\rm M}+D_{\rm m}$.  The fabrication and characterization of high-quality microtoroids are described in detail elsewhere~\cite{Armani:2003}.

The axisymmetric cavity modes of interest are whispering-gallery modes (WGM), which lie near the edge of the resonator surface and circulate in either a clockwise or counter-clockwise direction.  These modes are characterized by an azimuthal mode number $m$, whose magnitude gives the periodicity around the toroid and whose sign indicates the direction of propagation.  The WGMs for $\pm m$ are degenerate in frequency but travel around the toroid in opposite directions.  The mode electric fields for the WGM traveling waves are written as $\vect{E}(\vect{r}) = E_{\rm max} \vect{f}(\rho, z) e^{i m \phi}$, where $\vect{f}(\rho, z) = \vect{E}(\rho, z)/E_{\rm max}$ is the mode function in the $\rho-z$ cross-section normalized by the maximum electric field $E_{\rm max}$.  In general, backscattering couples these two modes so that a more useful eigenbasis for the system consists of the normal, standing wave modes characterized by a phase and the periodicity $|m|$.  This backscattering coupling $h$ is assumed to be real, with the phase absorbed into the definition of the origin of the coordinate $\phi$.  In addition, the mode's field decays at a rate $\kappa_{\rm i}$ through radiation, scattering, and absorption.  In our simulations, a cavity mode is fully described by its spatial mode function $\vect{f}(\vect{r})$, its azimuthal mode number $m$, its loss rate $\kappa_{\rm i}$, and the coupling $h$ to the counter-propagating mode with mode number $-m$.

We model the microtoroid modes using a commercial finite-element software package (COMSOL) to solve numerically for the vector mode functions $\vect{f}(\rho, z)$ for modes of a given $m$~\cite{Oxborrow:2007}.    Mode volumes are calculated from,
\begin{equation}
V_{\rm m} = \frac{\int \! dV \, \epsilon(\vect{r}) |\vect{E}(\vect{r})|^2}{E_{\rm max}^2} = 2 \pi \int \! dA \, \epsilon(\rho, z) \rho  f(\rho, z)^2
\end{equation}
In this notation~\cite{Spillane:2005}, the coupling of a circularly polarized optical field to an atomic dipole located in the evanescent field of the cavity is calculated as:
\begin{equation}
g(\vect{r}) = \langle \vect{d}\cdot \vect{E}\rangle = f(\rho, z) e^{i m \phi} \sqrt{\frac{3 \pi c^3 \gamma}{{\omega^{(0)} _{\rm a}}^2 V_{\rm m}}}
\end{equation}
where $\vect{d}$ is the dipole operator and $\omega^{(0)}_{\rm a} = 2\pi c/\lambda_0$ is the vacuum transition frequency of the two-level atom with free-space wavelength $\lambda_0$.  WGMs are predominantly linearly polarized, and so we average over the dipole matrix elements to obtain an effective traveling wave coupling $g_{\rm tw}$ which is approximately $\sim 0.6$ of the value for circularly polarized light~(see supplementary information of~\cite{Aoki:2006}.   Travelling wave modes of an axisymmetric resonator are not strictly transverse.  For the toroid geometries considered here, with $D_{\rm p}, D_{\rm m} \gg \lambda$, the azimuthal component is small and we assume that the optical field is linear outside of the toroid.  Since the cavity losses are dominated by absorption and defect scattering rather than the radiative lifetime set by the toroid geometry~\cite{Spillane:2005}, we let $\kappa_{\rm i}$ and $h$ be experimental parameters.   Fig.~\ref{Fig-modes} shows the lowest-order mode with $m=118$ for a toroid with $\{D_{\rm p},D_{\rm m}\} = \{24, 3\}$ $\mu$m. The index $m$ is chosen so that the cavity frequency $\omega_{\rm c}$ is near the $6S_{1/2}\rightarrow 6P_{3/2}$ transition of Cs with $\omega_{\rm a}^{(0)}/2\pi = 351.7$ THz.

The local polarization of modes varies throughout the interior and exterior of the toroid.  Approximate solutions for constant polarization suggest classifications as quasi-transverse modes, labeled transverse electric (TE) and transverse magnetic (TM) modes, although actual solutions are not transverse.  A reasonable analytic approximation for the lowest-order mode function with mode number $m$ outside of the toroid is that of a Gaussian wrapped around the toroid's surface that decays exponentially with distance scale set by the free space wavevector $1/\lambdabar_0 = 2 \pi/\lambda_0$,
\begin{equation} \label{eq:approxf}
f(\rho, z)\sim e^{-d/\lambdabar_0} e^{-(\psi/\psi_0)^2}
\end{equation}
where $d(\rho,z) = \sqrt{(\rho-D_{\rm M}/2)^2+z^2} - D_{\rm m}/2$ is the distance to the toroid surface, $\psi(\rho, z) = \arctan \frac{z}{\rho-D_{\rm M}/2}$ is the angle around the toroid cross-section ($\psi=0$ at $z=0$), and $\psi_0$ is a characteristic mode width (see Fig. 1a).  Higher order angular modes are characterized by additional nodes along the coordinate $\psi$.

\subsection{Cavity QED in an axisymmetric resonator}\label{sec:theoryequations}

\begin{figure}[tpb]
  \begin{center}
    \includegraphics{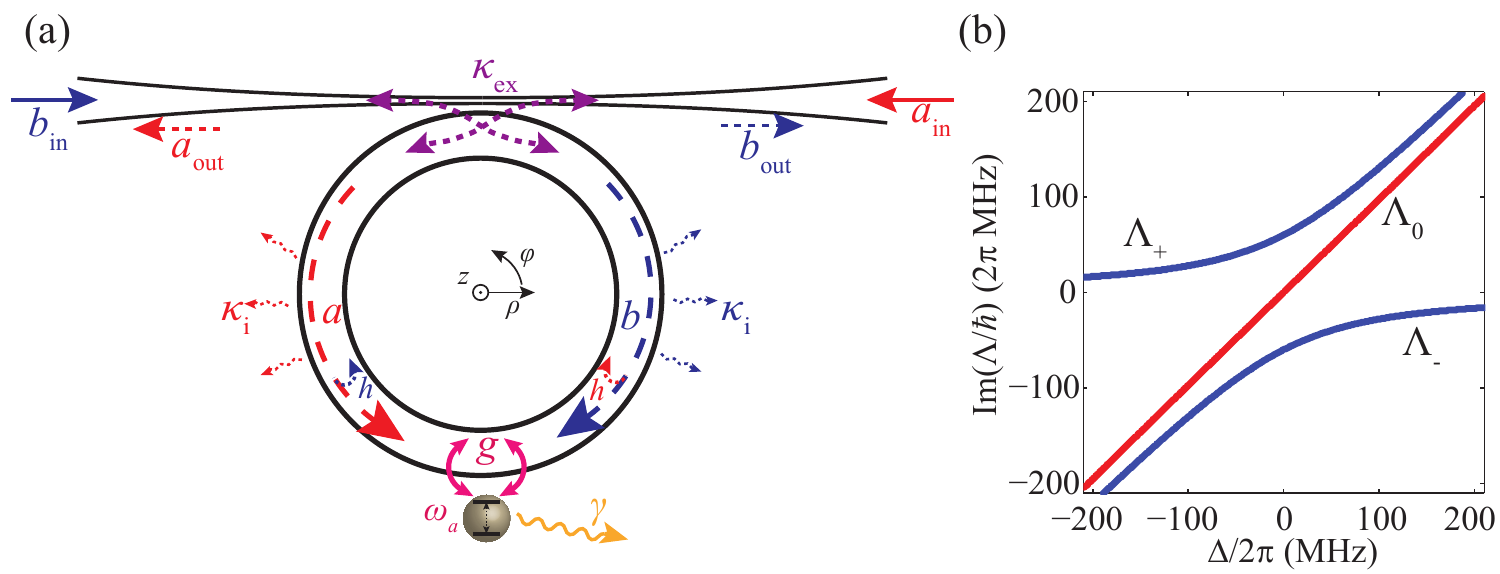}
  \end{center}\vspace{-0.5cm}
\caption{\label{Fig-Schematic} (a) Schematic of the atom-toroid system.  Coherent optical fields in the tapered fiber couple into whispering-gallery cavity modes of an axisymmetric resonator.  These fields can couple to an atomic transition with rate $g$, scatter to the counter-propagating mode ($h$), escape to the environment ($\kappa_{\rm i}$), or couple in/out of the fiber ($\kappa_{\rm ex}$).  The atom is described as a two-level system with transition frequency $\omega_{\rm a}$ and spontaneous emission rate $\gamma$.  (b) Imaginary part of the eigenvalues $\Lambda_i$ of the linearized systems as a function of detuning $\Delta= \omega_{\rm c}-\omega_{\rm a}^{(0)}$ for a Cs atom at $\phi = \pi/4$ and $g = 60$ MHz critically coupled to a cavity with parameters $\{\kappa_{\rm i}, h\}/ 2 \pi = \{8, 0\} $ MHz (Eqs.~\eref{Eq:lineqs}).}
\end{figure}

We consider a quantum model of a two-level atom at position $\vect{r}(t)$ coupled to an axisymmetric resonator shown schematically in Fig.~\ref{Fig-Schematic}.  The terminology used here follows the supplemental material of~\cite{Aoki:2006},~\cite{Dayan:2008}, and~\cite{Alton:2010}, but the general formalism can be found in additional sources (see \cite{Srinivasan:2007}, for example).  As described in section~\ref{subsec:modes}, an axisymmetric resonator supports two degenerate counter-propagating whispering-gallery modes at resonance frequency $\omega_{\rm c}$ to which we associate the annihilation (creation) operators $a$ and $b$ ($a^\dagger$ and $b^\dagger$).  Each traveling-wave mode has an intrinsic loss rate, $\kappa_{\rm i}$; the modes are coupled via scattering at rate $h$.  External optical access to the cavity is provided by a tapered fiber carrying input fields $\{a_{\rm in}, b_{\rm in} \}$ at probe frequency $\omega_{\rm p}$.  Fiber fields couple to the cavity modes with an external coupling rate $\kappa_{\rm ex}$.  The output fields of the fiber taper in each direction are the coherent sum of the input field and the leaking cavity field, $\{a_{\rm out}, b_{\rm out}\} = -\{a_{\rm in},  b_{\rm in}\} + \sqrt{2\kappa_{\rm ex}} \{a, b\}$~\cite{Aoki:2006, Dayan:2008}.

We specialize to the situation of single-sided excitation, where $\langle b_{\rm in}\rangle=0$.  The input field $a_{\rm in}$ drives the $a$ mode with strength $\varepsilon_{\rm p} = i \sqrt{2 \kappa_{\rm ex}} \langle a_{\rm in} \rangle$ so that the incident photon flux is $P_{\rm in}=\langle a^\dagger_{\rm in} a_{\rm in}\rangle = |\varepsilon_{\rm p}|^2/2\kappa_{\rm ex}$.  Experimentally accessible quantities are the transmitted and reflected photon fluxes, $P_{\rm T}=\langle a^\dagger_{\rm out} a_{\rm out}\rangle$ and $P_{\rm R}=\langle b^\dagger_{\rm out} b_{\rm out}\rangle$, respectively.  In experiments, data is typically presented as normalized transmission and reflection coefficients defined as $T = P_{\rm T}/P_{\rm in}$ and $R = P_{\rm R}/P_{\rm in}$.  In the absence of an atom, the functions $T$ and $R$ for the bare cavity depend on the detuning $\Delta_{\rm cp} = \omega_{\rm c} - \omega_{\rm p}$ and the cavity rates $h$, $\kappa_{\rm i}$, and $\kappa_{\rm ex}$.  At critical coupling, $\kappa_{\rm ex} = \sqrt{\kappa_{\rm i}^2 + h^2}$, the bare cavity $T\rightarrow 0$ when $\Delta_{\rm cp} = 0$~\cite{Spillane:2003}.

The cavity modes $\{a, b\}$ both couple to a single two-level atom with transition frequency $\omega_{\rm a}$ at location $\vect{r}$.  In the context of cQED, the atomic system is described by a single transition with frequency $\omega_{\rm a}$ with the associated raising and lowering operators $\sigma^{+}$ and $\sigma^{-}$ and an excited state field decay rate $\gamma$. The atomic frequency $\omega_{\rm a}(\vect{r})$ may be shifted from the free-space value $\omega^{(0)}_{\rm a}$ by frequency $\delta_{\rm a}(\vect{r})$ due to interactions with the dielectric surface.  The coupling of the traveling-wave modes $\{a,b\}$ to the atomic dipole is described by the single-photon coupling rate $g_{\rm tw}(\vect{r})=g_{\rm tw}^{\rm max} f(\rho,z) {\rm e}^{\pm i \theta}$, where $f(\rho,z)$ is the cavity mode function and $\theta = m \phi$.  A discussion of $f(\rho,z)$ for the modes of a microtoroid appears in section~\ref{subsec:modes}.  For an atom in motion, $\omega_{\rm a}(\vect{r})$, $\gamma(\vect{r})$, and $g_{\rm tw}(\vect{r})$ are spatially-dependent quantities that depend on the atomic position $\vect{r}(t)$.

To study the atom-cavity dynamics, we write the standard Jaynes-Cummings-style cQED Hamiltonian for coupled field modes~\cite{JaynesCummings:1963, Aoki:2006}:
\begin{eqnarray}
H/\hbar =&\omega_{\rm a}(\vect{r}) \sigma^+\sigma^- + \omega_{\rm c}\left( a^\dagger a + b^\dagger b \right)  + h \left( a^\dagger b + b^\dagger a \right) +\left(\varepsilon^*_{\rm p}e^{i \omega_{\rm p}t} a + \varepsilon_{\rm p}e^{-i \omega_{\rm p}t} a^\dagger \right)\nonumber\\ \nopagebreak
  &+\left(g^*_{\rm tw}(\vect{r}) a^\dagger\sigma^- + g_{\rm tw}(\vect{r}) \sigma^+ a \right) + \left( g_{\rm tw}(\vect{r}) b^\dagger \sigma^- + g^*_{\rm tw}(\vect{r}) \sigma^+ b \right).
\end{eqnarray}
Following the rotating-wave approximation, we write the Hamiltonian in a frame rotating at $\omega_{\rm p}$~\cite{Aoki:2006, Dayan:2008, Srinivasan:2007}:
\begin{eqnarray}\label{eq:Hamiltonian1}
H/\hbar =&\Delta_{\rm ap}(\vect{r})\sigma^+\sigma^- + \Delta_{\rm cp}\left( a^\dagger a + b^\dagger b \right)  + h \left( a^\dagger b + b^\dagger a \right) +\varepsilon^*_{\rm p} a + \varepsilon_{\rm p} a^\dagger \nonumber \\ &+ \left(g^*_{\rm tw}(\vect{r}) a^\dagger\sigma^- + g_{\rm tw}(\vect{r}) \sigma^+ a \right) + \left( g_{\rm tw}(\vect{r}) b^\dagger \sigma^- + g^*_{\rm tw}(\vect{r}) \sigma^+ b \right),
\end{eqnarray}
where $\Delta_{\rm ap}(\vect{r}) = \omega_{\rm a}(\vect{r}) - \omega_{\rm p}$ and $\Delta_{\rm cp} = \omega_{\rm c} - \omega_{\rm p}$.  Dissipation from coupling to external modes is treated using the master equation for the density operator of the system $\rho$:
\begin{eqnarray}\label{eq:MasterEquation}
\dot{\rho}=&-\frac{i}{\hbar} [H,\rho]+\kappa\left(2a\rho a^\dagger-a^\dagger a\rho - \rho a^\dagger a\right) + \kappa \left( 2b\rho b^\dagger - b^\dagger b \rho - \rho b^\dagger b\right)\nonumber \\ &+ \gamma\left(2\sigma^-\rho\sigma^+ - \sigma^+\sigma^-\rho - \rho\sigma^+\sigma^-\right)
\end{eqnarray}
Here, $\kappa = \kappa_{\rm i} + \kappa_{\rm ex}$ is the total field decay rate of each cavity mode, and $2 \gamma(\vect{r})$ is the atomic dipole spontaneous emission rate, which is orientation dependent near a dielectric surface (section~\ref{sec:SpontEm}).

The Hamiltonian~\eref{eq:Hamiltonian1} can be rewritten in a standing wave basis using normal modes $A = (a+b)/\sqrt{2}$ and $B = (a-b)/\sqrt{2}$,
\begin{eqnarray}\label{eq:Hamiltonian2}
H/\hbar = &\Delta_{\rm ap}(\vect{r}) \sigma^+\sigma^- + (\Delta_{\rm cp}+h) A^\dagger A + (\Delta_{\rm cp}-h) B^\dagger B \nonumber\\&+  \left( \varepsilon^*_{\rm p} A + \varepsilon_{\rm p} A^\dagger \right)/\sqrt{2} +  \left( \varepsilon^*_{\rm p} B + \varepsilon_{\rm p} B^\dagger \right)/\sqrt{2} \nonumber  \\
&+ g_{\rm A}(\vect{r}) \left( A^\dagger\sigma^- + \sigma^+ A \right) - {i} g_{\rm B}(\vect{r}) \left( B^\dagger\sigma^- - \sigma^+ B \right),
\end{eqnarray}
where $g_{\rm A}(\vect{r}) = g_{\rm max} f(\rho,z) \cos \theta$, $g_{\rm B}(\vect{r}) = g_{\rm max} f(\rho,z) \sin\theta$, and $g_{\rm max} = \sqrt{2} g_{\rm tw}^{\rm max}$. In the absence of atomic coupling ($g_{\rm tw} = 0$), these normal modes are eigenstates of~\eref{eq:Hamiltonian1}.  With $g_{\rm tw} \neq 0$, the eigenstates of the Hamiltonian are dressed states of atom-cavity excitations.  With $h=0$ and $g_{\rm tw} \neq 0$, the atom defines a natural basis in which it couples to only a single standing wave mode.  For the modes $\{A, B\}$ defined above, coupling may occur predominantly, or even exclusively, to one of the two normal modes depending the azimuthal coordinate $\theta$. For such $\theta$, the system can be interpreted as an atom coupled to one normal mode in a traditional Jaynes-Cummings model with dressed-state splitting given by the single-photon Rabi frequency $\Omega_{(1)} = 2 g \equiv 2 g_{\rm max} f(\rho,z)$, along with a second complementary cavity mode uncoupled to the atom.  Approximately for $g_{\rm tw} \gg h$, this interpretation is consistent for any arbitrary atomic coordinate $\theta$. For $h \neq 0 $ and comparable to $\kappa_{\rm i}$ with a fixed phase convention (such as Im$(h)=0$ used here), this decomposition is not possible for arbitrary atomic coordinate $\theta$; the atom in general couples to both normal modes as a function of $\phi$~\cite{Aoki:2006, Srinivasan:2007}.

The master equation~\eref{eq:MasterEquation} can be numerically solved using a truncated number state basis for the cavity modes. Alternatively, the system is linearized by treating the atom operators $\sigma^{\pm}$ as approximate bosonic harmonic oscillator operators with $\left[\sigma^-,\sigma^+\right] \approx 1$. For a sufficiently weak probe field, the atomic excited state population is small enough that the oscillator has negligible population above the first excited level and the harmonic approximation is quite good.  As part of this linearization, we factor expectation values of normally-ordered operator products into products of operator expectation values~\cite{Doherty:1997, Fischer:2001}.  Reducing operators to complex numbers suppresses coherence, but numerical calculations confirm that this approximation is accurate when calculating cavity output fields and classical forces for the weak driving power levels considered here.  In particular, experiments typically utilize a photon flux $P_{\rm T}=\langle a^\dagger_{\rm out} a_{\rm out}\rangle \sim 15$ cts/$\mu$s corresponding to an average cavity photon population of $\langle a^\dagger a \rangle \sim 0.1$.  At these photon numbers, cavity expectation values effectively factorize such that $\langle a^\dagger a \rangle \approx \langle a^\dagger \rangle \langle a \rangle$ for the semiclassical treatment used here~\cite{Hood:1998}.  We use this approximation to write $P_{\rm T}=\langle a^\dagger_{\rm out} a_{\rm out}\rangle \approx  \langle a^\dagger_{\rm out} \rangle \langle a_{\rm out}\rangle$, implying that we only need the complex number $\langle a_{\rm out}\rangle =-\langle a_{\rm in}\rangle + \sqrt{2\kappa_{\rm ex}} \langle a \rangle$  and its conjugate to calculate the cavity transmission at these photon numbers.  This approximation is not sufficient for calculation of the $g^{(2)}(\tau)$ correlation function where the nonlinearities must be included~\cite{Alton:2010}.

The relevant equations of motion for the field amplitudes of the linearized master equation are,
\numparts
\begin{eqnarray}\label{Eq:lineqs}
\langle \dot{a} \rangle & =  -(\kappa + {i} \Delta_{\rm cp})~\langle a \rangle - {i} h~\langle b \rangle - {i} \varepsilon_{\rm p} - {i} g^*_{\rm tw}~\langle \sigma^- \rangle,  \\
\langle \dot{b} \rangle & =  -(\kappa + {i} \Delta_{\rm cp} )~\langle b \rangle - {i} h~\langle a \rangle - {i} g_{\rm tw}~\langle \sigma^- \rangle,  \\
\langle \dot{\sigma^-} \rangle & =  -(\gamma + {i} \Delta_{\rm ap} )~\langle \sigma^- \rangle - {i} g_{\rm tw}~\langle a \rangle - {i} g^*_{\rm tw}~\langle b \rangle.  \label{Eq:sigma}
\end{eqnarray}
\endnumparts

Time and spectral dependence of this system of equations are governed by its eigenvalues $\Lambda_i$.  The imaginary part of the eigenvalues as a function of detuning $\Delta \equiv \Delta_{\rm cp} - \Delta_{\rm ap} = \omega_{\rm c}-\omega_{\rm a}$ are illustrated in Fig.~\ref{Fig-Schematic}b. For large $\Delta\gg |g_{\rm tw}|$, the three eigenvalues include one atom-like eigenvalue and two cavity-like eigenvalues separated by the mode splitting $h$.  For intermediate $\Delta$, there is an anti-crossing of two dressed-state eigenvalues $\Lambda_{\pm}$, while the third (cavity-like) $\Lambda_0$ is uncoupled to the atom.

For a slowly-moving atom, the mode fields remain approximately in steady state as the parameters evolve with the atom trajectory $\vect{r}(t)$.  Analytic steady-state solutions to~\eref{Eq:lineqs} for $\langle a \rangle_{\rm ss}$ and $\langle b \rangle_{\rm ss}$ are:

\begin{small}
\numparts
\begin{eqnarray}\label{eq:linsols}
\hspace{-3.0cm}\langle a \rangle_{\rm ss} & =  {i} \varepsilon_{\rm p} \frac{\left(\gamma + {i}\Delta_{\rm ap}\right)\left[(\kappa + {i}\Delta_{\rm cp})\left(\gamma+{i}\Delta_{\rm ap}\right)+|g_{\rm tw}|^2\right]}{\left[{i} h (\gamma + {i} \Delta_{\rm ap})+(g^*_{\rm tw})^2 \right] \left[{i} h (\gamma+{i}\Delta_{\rm ap})+g_{\rm tw}^2\right]-\left[(\kappa+{i}\Delta_{\rm cp})(\gamma+{i}\Delta_{\rm ap})+|g_{\rm tw}|^2\right]^2} \label{eq:linsolsa} \\
\hspace{-3.0cm}\langle b \rangle_{\rm ss} & =  - \frac{{i} h \left(\gamma+{i}\Delta_{\rm ap}\right)+g^2_{\rm tw}}{(\kappa+{i}\Delta_{\rm cp})(\gamma+{i}\Delta_{\rm ap})+|g_{\rm tw}|^2} \langle a \rangle_{\rm ss} \label{eq:linsolsb} \\
\hspace{-3.0cm}\langle \sigma^- \rangle_{\rm ss} & =  -i \frac{g_{\rm tw} \langle a\rangle_{\rm ss} + g^*_{\rm tw} \langle b\rangle_{\rm ss} }{\gamma+i \Delta_{\rm ap}}. \label{eq:linsolss}
\end{eqnarray}
\end{small}
\endnumparts

\section{Optical forces on an atom in a cavity}\label{sec:OpticalForce}

Neutral atoms experience forces from the interaction of the atomic dipole moment with the radiation field.  These optical dipole forces have a quantum mechanical interpretation as coherent photon scattering~\cite{Gordon:1980, Dalibard:1985}.  For a light field near resonance with the atomic dipole transition, these optical forces can be quite strong, even at the single photon level; cavity-enhanced dipole forces~\cite{Doherty:1997, Horak:1997} have been exploited to trap~\cite{Hood:2000} and localize~\cite{Pinkse:2000} a single atom with the force generated by a single strongly-coupled photon.  In this section, we discuss how the optical forces, their first-order velocity dependence, and their fluctuations are included in our semiclassical simulation.

\subsection{Dipole forces}

In a quantum mechanical treatment of light-matter interactions~\cite{Dalibard:1985}, the eigenstates of the system are dressed states of atom  and optical field.  The quantum mechanical optical force on the atom at location $\vect{r}$ can be found from the commutator of the atom momentum $\vect{p}$ with the interaction Hamiltonian $H_{\rm int}$ consisting of the last two terms from the Hamiltonian~\eref{eq:Hamiltonian1}:
\begin{equation}\label{Eq:force}
\hspace{-2cm}\vect{F} = \frac{d\vect{p}}{d t} = \frac{i}{\hbar}\left[H_{\rm int}, \vect{p} \right]  = -  \hbar \nabla g^*_{\rm tw}(\vect{r}) \left(a^\dagger \sigma^- + \sigma^+ b\right) - \hbar \nabla g_{\rm tw}(\vect{r}) \left( \sigma^+ a + b^\dagger \sigma^-\right)
\end{equation}
The gradient from the position space representation of the momentum operator $\vect{p}$ only acts on $g_{\rm tw}(\vect{r})$ and not on the field operators~\cite{vanEnk:2001, Domokos:2002}.  The steady-state expectation values of~\eref{Eq:force} give the dipole force on the atom in the semiclassical approximation:
\begin{eqnarray}\label{Eq:forceSC}
\hspace{-2cm}\langle\vect{F}\rangle_{\rm ss} =  &-  \hbar \nabla g^*_{\rm tw}(\vect{r}) \left(\langle a^\dagger\rangle_{\rm ss} \langle\sigma^-\rangle_{\rm ss} + \langle\sigma^+\rangle_{\rm ss} \langle b \rangle_{\rm ss}\right) \nonumber \\ &- \hbar \nabla g_{\rm tw}(\vect{r}) \left( \langle \sigma^+\rangle_{\rm ss} \langle a\rangle_{\rm ss} + \langle b^\dagger\rangle_{\rm ss} \langle\sigma^-\rangle_{\rm ss}\right)
\end{eqnarray}
As described in section~\ref{Sec:theory}, the steady-state operator expressions are simplified by reducing expectation values of operator products to products of linearized steady-state operator expectation values.  Ignoring fiber and spontaneous emission losses, an effective conservative dipole potential $U_{\rm d}$ can be defined by integration of~\eref{Eq:forceSC}.

\subsection{Velocity-dependent forces on an atom}

Non-zero velocity effects on the force~\eref{Eq:forceSC} are found by including a first-order velocity correction in the steady state expectation values~\cite{ Fischer:2001, Gordon:1980, Domokos:2002}. Consider a vector of operators $\vect{O}$ whose expectation values obey a linearized equation system such as~\eref{Eq:lineqs}. Assuming a small velocity, we expand the operator expectation values $\langle \vect{O} \rangle$ as:
\begin{equation}\label{eq:vdep}
\langle \vect{O} \rangle = \langle \vect{O} \rangle_0 + \langle \vect{O} \rangle_1 + \ldots,
\end{equation}
where the subscripts denote the order of the velocity $v$ in each term.  If an atom is moving through these fields, then the cavity parameters depend in general on atomic position $\vect{r}$.  As $\vect{r}$ changes in time, the fields must evolve in response.  Consequently, the time derivative of the expectation value evolves not only from explicit time dependence, but from atomic motion as well.
\begin{equation}\label{eq:vdepss}
\langle \dot{\vect{O}} \rangle = \left(\frac{\partial}{\partial t} + \vect{v}\cdot \vect{\nabla} \right) \langle \vect{O} \rangle
\end{equation}
Setting the explicit time derivatives in~\eref{eq:vdepss} to zero, the perturbative expansion of the time derivative can be equated to the original linearized equation system.  Collecting terms of each order in velocity gives an equation for the first-order term $\langle \vect{O} \rangle_1$ in terms of the zero-velocity steady-state solution $\langle \vect{O} \rangle_0$.  This procedure requires the spatial derivative of the zero-order steady-state solutions, where spatial dependence enters through the atomic transition frequency $\omega_{\rm a}(\vect{r})$, the spontaneous emission rate $\gamma(\vect{r})$, and the atom-cavity coupling $g(\vect{r})$.  Only terms linear in velocity are kept in the operator products of the force $\vect{F}(\vect{r})$ in~\eref{Eq:forceSC}.

In practice, first-order velocity corrections are small in our simulation. For example, Doppler shifts arising from spatial derivatives of the cavity modes are on the order of $\vect{k}\cdot \vect{v}$, where $\vect{k}$ is the mode wavevector.  For typical azimuthal velocities of less than 0.1 m/s, the Doppler shift is less than 1 MHz.  The effect becomes more significant as atoms accelerate to high velocities near the surface, but atomic level shifts from surface interactions are more significant in this regime than the Doppler shifts. 

\subsection{Momentum diffusion and the diffusion tensor in a cavity}

Quantum fluctuations of optical forces are treated by adding a stochastic momentum diffusion contribution to the atomic velocity in the Langevin equations of motion. We calculate the diffusion tensor components used in~\eref{eq:langevin}, $D_{ii}$, using general expressions for diffusion in an atom-cavity system generalized for the two-mode cavity of a toroid~\cite{Murr:2006}:
\begin{equation}\label{Eq:Diffusion}
\hspace{-2cm}2 D_{ii} = (\hbar k)^2 2\gamma \left|\left\langle \sigma^- \right\rangle_{\rm ss}\right|^2 + \left| \hbar \nabla_i \left\langle \sigma^- \right\rangle_{\rm ss} \right|^2 2 \gamma + 2\kappa \left(\left| \hbar \nabla_i \left\langle a \right\rangle_{\rm ss} \right|^2 + \left| \hbar \nabla_i \left\langle b \right\rangle_{\rm ss} \right|^2\right)
\end{equation}
for $i=x,y,z$, where $\gamma$ is the atomic field spontaneous decay rate. The first term represents fluctuations from spontaneous emission, the second term describes a fluctuating atomic dipole coupled to a cavity field, and the third represents a fluctuating cavity field coupled to an atomic dipole.~\eref{Eq:Diffusion} is approximated using steady-state fields calculated from the linearized solutions to the master equation~\eref{eq:linsols}.  Although included in the trajectory model, momentum diffusion does not significantly alter averages over ensembles of trajectories at the weak excitation levels and low atomic velocities used in the relevant experiments.

\section{Effects of surfaces on atoms near dielectrics}\label{sec:Surface}

In the vicinity of a material surface, the mode structure of the full electromagnetic field is modified due to the dielectric properties of nearby objects.  These off-resonant radiative interactions modify the dipole decay rate of atomic states and shift electronic energy levels.  This surface interaction varies spatially as the relative atom-surface configuration changes.  The surface phenomena are dispersive and depend on the multi-level description of the atom's electronic structure; they are calculated using traditional perturbation theory with the full electromagnetic field without focusing on a few select modes enhanced by a cavity in cQED.

\subsection{Spontaneous emission rate near a surface}\label{sec:SpontEm}

When a classical oscillating dipole is placed near a surface, its radiation pattern is modified by the time-lagged reflected field from the dielectric surface.  The spontaneous emission rate oscillates with distance $d$ from the surface, which can be interpreted as interference between the radiation field of the dipole and its reflection.  The variation of the emission rate depends on whether the dipole vector is parallel or perpendicular to the surface, as intuitively expected from the asymmetry of image dipole orientations of dipoles aligned parallel and perpendicular to the surface normal.  For either orientation, the spontaneous emission rate features a marked increase within a wavelength of the surface due to surface evanescent modes that become available for decay for $d \lesssim \lambdabar_0$.

\begin{figure}[tpb]
  \begin{center}
    \includegraphics{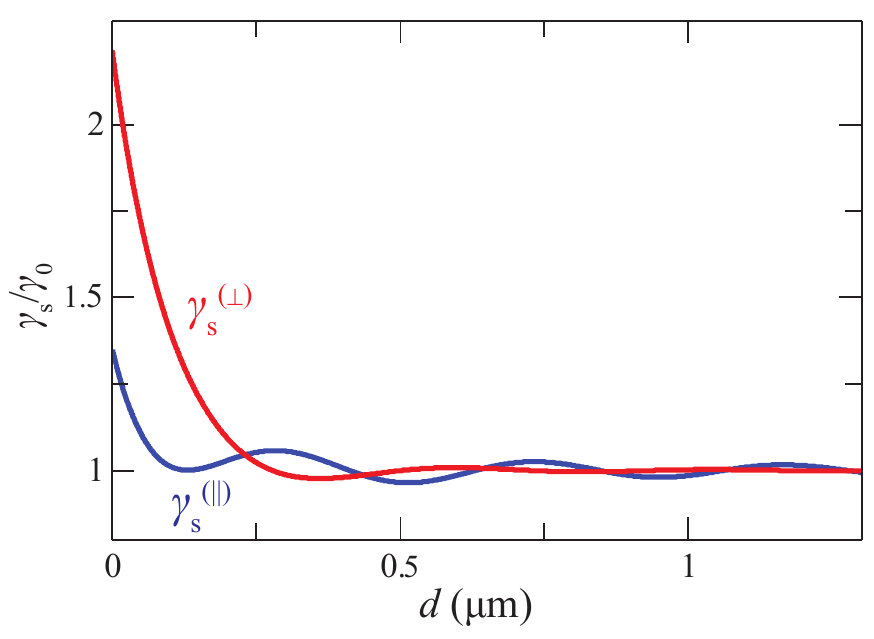}
  \end{center}\vspace{-0.5cm}
\caption{Variations of the dipole decay rate $\gamma_{\rm s}(d)$ for a dipole oriented parallel $(\|)$ and perpendicular $(\bot)$ to the surface normal as a function of distance $d$ from a semi-infinite region of SiO$_2$. The decay rate is in units of the vacuum decay rate $\gamma_0$ and the wavelength of the transition is $\lambda = 852$ nm.} \label{fig-gammadistance}
\end{figure}

We calculate the surface-modified dipole decay rates $\gamma^{(\|)}_{\rm s}(d)$ and $\gamma^{(\bot)}_{\rm s}(d)$ for a cesium atom near an SiO$_2$ surface following the methods of Refs.~\cite{Lukosz:1977, Li:1990} (see Fig.~\ref{fig-gammadistance}).  This calculation involves an integration of surface reflection coefficients over possible wavevectors of radiated light.  The integrand depends on the dielectric function of SiO$_2$ evaluated at the frequency $\omega_{\rm a}$ of the atomic transition.  The orientations refer to the alignment of a classical dipole relative to the surface plane.

\subsection{Calculation of Casimir-Polder potentials}

Radiative interactions with a surface are important components of motion for neutral atoms within a few hundred nm of a surface, with the potential for manipulating atomic motion through attractive~\cite{Sukenik:1993} or repulsive forces~\cite{Munday:2009}.  Depending on the theoretical framework, these forces are naturally thought of as radiative self-interactions between two polarizable objects, fluctuations of virtual electromagnetic excitations, or as a manifestation of vacuum energy of the electromagnetic field.  These surface interactions, represented by a conservative potential $U_{\rm s}$, are sensitive to the frequency dispersion of the electromagnetic response properties of the atoms and surfaces.

For an atom located a short distance $d\ll \lambda_0$ from a dielectric, the fluctuating dipole of the atom interacts with its own surface image dipole in the well-known nonretarded van der Waals interaction.  Using only classical electrodynamics with a fluctuating dipole, the surface interaction potential is found to take the Lennard-Jones (LJ) form $U_{\rm s}^{\rm LJ} = -C_3/d^3$, where $C_3$ is a constant that depends on the atomic polarizability and dielectric permittivity of the surface~\cite{London:1930, London:1937, Lennard-Jones:1932, Fichet:1995}.  At larger separations, virtual photons exchanged between atoms and surfaces cannot travel the distance in time $t \sim 1/\omega$ due to the finite speed of light.  Consequently, the interaction potential is reduced, as first calculated in the 1948 paper by Casimir and Polder \cite{Casimir:1948}.  The retarded surface potential takes the form $U_{\rm s}^{\rm ret} = -C_4/d^4$ for a constant $C_4$, where $C_4$ depends on both $c$ and $\hbar$ as this is fundamentally both a relativistic and quantum phenomenon.  The full theory of surface forces for real materials with dispersive dielectric functions came with the work of Lifshitz~\cite{Lifshitz:1956, Dzyaloshinskii:1961}. This framework reduces to both the above situations for the proper limits, and, importantly, it accounts for finite temperatures, predicting a $U_{\rm s}^{\rm th} \propto d^{-3}$ potential caused by thermal photons dominant at large distances for $d \gg \hbar c/k_B T$~\cite{Antezza:2004}.  In our discussion, we refer to these generalized dispersion forces as \emph{Casimir-Polder} (CP) forces, whereas $U_{\rm s}^{\rm LJ} $, $U_{\rm s}^{\rm ret}$, and $U_{\rm s}^{\rm th}$ refer to the appropriate distance limits.

In microcavity cQED, evanescent field distance scales are set by the scale length of the evanescent field, $\lambdabar_0 = \lambda_0/2\pi = 136$ nm (for the Cs $D2$ line). The relevant distances ($0 < d \lesssim 300$ nm) span both the LJ and retarded regimes, but are much shorter than the thermal regime ($d>~5~\mu$m). In the transition region, the limiting power laws do not fully describe $U_{\rm s}$ over the relevant range of $d$. In our modeling, we utilize a calculation of $U_{\rm s}$ with the Lifshitz approach.   The Lennard-Jones, retarded, and thermal limits arise naturally from the Lifshitz formalism~\cite{Antezza:2004}.

The potential $U_{\rm s}$ enters our simulation in two ways.   First, the transition frequency $\omega_{\rm a}$ of the two-level atomic system shifts away from the vacuum frequency by $\delta_{\rm a} = (U_{\rm s}^{\rm ex}(\vect{r}) - U_{\rm s}^{\rm g}(\vect{r}))/\hbar$, where $U_{\rm s}^{\rm g}(\vect{r})$ and $U_{\rm s}^{\rm ex}(\vect{r})$ are the surface potentials for the ground and excited states, respectively.  Since the atom transitions between the ground and excited states during its passage through the mode, the average net force used in calculations is found by weighting each contribution by the steady-state atomic state populations, $F_{\rm s} = F_{\rm s}^{\rm g} \left( 1- \langle \sigma^{\dag} \rangle_{\rm ss} \left\langle \sigma \right\rangle_{\rm ss } \right) + F_{\rm s}^{\rm ex} \langle \sigma^{\dag} \rangle_{\rm ss } \left\langle \sigma \right\rangle_{\rm ss }$.

\begin{figure}[tpb]
  \begin{center}
    \includegraphics{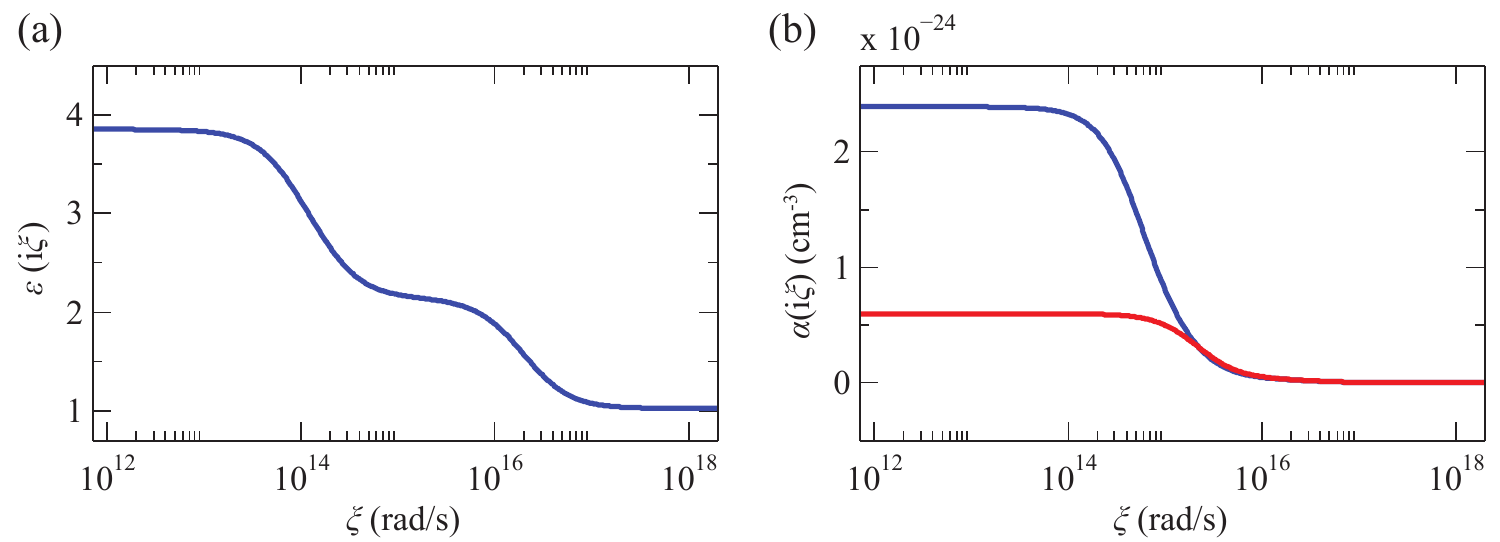}
  \end{center}\vspace{-0.5cm}
\caption{\label{Fig-Response} Dispersive response functions for SiO$_2$ and Cesium atoms.  (a) The dielectric function $\epsilon (i \xi)$ for SiO$_2$ evaluated for frequency $\xi$ along the imaginary axis.  (b) Total atomic polarizability $\alpha (i \xi)$ for SiO$_2$ evaluated for frequency $\xi$ along the imaginary axis for the $6S_{1/2}$ ground state (red) and the $6P_{3/2}$ excited state calculated as described in~\ref{app:response}. }
\end{figure}

We calculate $U_{\rm s}^{\rm g}$ and $U_{\rm s}^{\rm ex}$ for a cesium atom near a glass SiO$_2$ surface using the Lifshitz approach.  This calculation depends on the dispersion properties of the response functions of materials, in this case the polarizability of the Cs ground state $\alpha(\omega)$ and the complex dielectric function $\epsilon(\omega)$ of the silica surface.  Modeling of these functions is discussed in~\ref{app:response}. In particular, these response functions must be evaluated on the imaginary frequency axis $\omega = i \xi$, as shown in Figure~\ref{Fig-Response}.

Following the method of~\cite{Blagov:2005}, curvature of the toroid surface is implemented by treating the toroid as a cylinder with radius of curvature $R = D_{\rm m}/2$.  The major axis curvature is neglected because for all relevant distances $d \ll D_M/2$.  The resulting formula can be interpreted as a sum over discrete Matsubara frequencies $\xi_n = 2 \pi n k_B T/\hbar$ with an integration over transverse wave vectors, which we quote without derivation:~\cite{Blagov:2005}
\begin{eqnarray}\label{Eq:CPLifshitz}
\hspace{-1cm}U_{\rm surf}(d) = &- k_B T \sqrt{\frac{R}{R + d}} {\sum_{n=0}^{\infty}}^{\prime} \alpha(i \xi_n) \int_0^{\infty} k_{\perp}  \,d k_{\perp} \,e^{- 2 q_n d} \left[ q_n - \frac{1}{4(R + d)} \right] \nonumber \\ \hspace{-1cm}&\left\{ 2 r_{\parallel}(i \xi_n, k_{\perp}) + \frac{\xi_n^2}{q_n^2 c^2} \left[r_{\perp}(i \xi_n, k_{\perp}) - r_{\parallel}(i \xi_n, k_{\perp}) \right] \right\}
\end{eqnarray}
Here, $\alpha\left(i \xi_n\right)$ is the atomic polarizability and $r_{\parallel, \perp}\left(i \xi_n, k_{\perp}\right)$ are the reflection coefficients of the dielectric material evaluated for imaginary frequency $i \xi_n$. The primed summation implies a factor of $1/2$ for the $n =0$ term.  The reflection coefficients for the two orthogonal light polarizations are:
\begin{eqnarray}\label{Eq:CPRef}
r_{\parallel}\left(i\xi_n, k_{\perp}\right) &= \frac{\epsilon\left(i \xi_n\right) q_n - k_n}{\epsilon\left(i \xi_n\right) q_n + k_n}\\
r_{\perp}\left(i\xi_n, k_{\perp}\right) &= \frac{k_n - q_n}{k_n + q_n}
\end{eqnarray}
where
\begin{equation}
q_n = \sqrt{k_{\perp}^2 + \frac{\xi_{n}^2}{c^2}},\qquad k_{n} = \sqrt{k_{\perp}^2 + \epsilon\frac{\xi^2_n}{c^2}}
\end{equation}
and $\epsilon(i \xi_{n})$ is the complex dielectric function evaluated for imaginary frequencies $i \xi_{n}$.  Depending on the author, $r_{\parallel}$ ($r_{\perp}$) is sometimes referred to as $r_{\rm TM}$ ($r_{\rm TE}$).

$U_{\rm s}^{\rm g}$ is calculated by numerical evaluation of~\eref{Eq:CPLifshitz}.  $U_{\rm s}^{\rm ex}$ is also calculated using~\eref{Eq:CPLifshitz}, but with an additional contribution accounting for real photon exchange from the excited state with the surface, which is proportional to ${\rm Re}\left[ \frac{\epsilon(\omega_{\rm a}) - 1}{\epsilon(\omega_{\rm a}) + 1}\right]$ in the LJ limit~\cite{Fichet:1995, Gorza:2006}.

\begin{figure}[tpb]
  \begin{center}
    \includegraphics{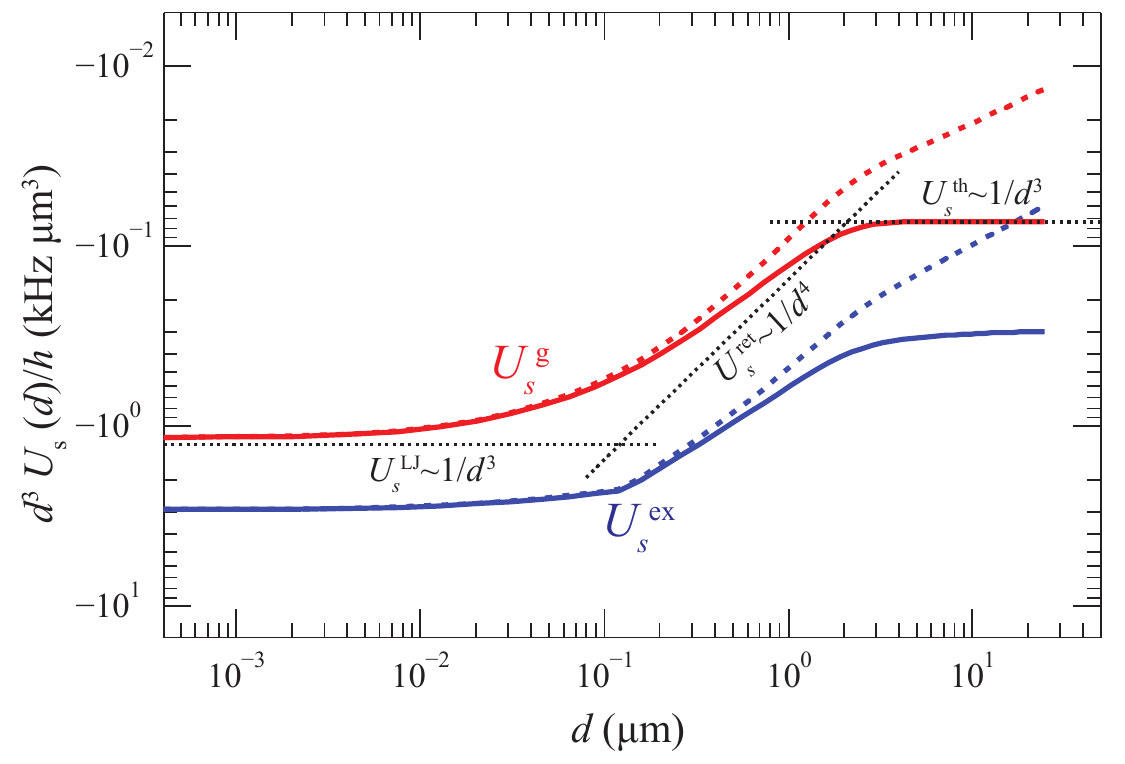}
  \end{center}\vspace{-0.5cm}
\caption{\label{Fig-CP} Atom-surface potentials $U_{\rm s}^{\rm g}$ (red) and $U_{\rm s}^{\rm ex}$ (blue) for a cesium atom at distance $d$ from a SiO$_2$ surface.  The solid lines are for a planar surface whereas the dashed lines are for a curved surface with radius of curvature $R = D_{\rm m}/2 = 1.5$ $\mu$m.  The limiting regimes for $U_{\rm s}^{\rm g}$ with a planar surface are shown as dotted lines, each calculated from analytic expressions not using the Lifshitz formalism.  The cylindrical surface correction weakens the potential, which is noticeable in the retarded and thermal regimes. }
\end{figure}

The atom-surface potential $U_{\rm s}^{\rm g}$ for the ground state of cesium near a SiO$_2$ surface is shown in Fig.~\ref{Fig-CP}, including calculations for both a planar and a cylindrical surface.  Without the cylindrical correction, the potential approaches the LJ, retarded, and thermal limits at appropriate distance scales. For the planar dielectric, our calculation yields $C_3/h = 1178$ Hz $\mu$m$^3$ and $C_4/h = 158$ Hz $\mu$m$^4$ for the LJ and retarded limits.  Note that the transition region between LJ and retarded regimes occurs around $d\sim100$ nm, the relevant distance scale for the experiments we are modeling.  For $d > D_{\rm m}$, the perturbative method accounting for the curvature is no longer accurate~\cite{Blagov:2005}, but at these distances, the surface forces are insignificant to atomic motion due to their steep power law fall-off.  The excited state potential $U_{\rm s}^{\rm ex}$ has a similar form to $U_{\rm s}^{\rm g}$.

\section{Simulating atoms detected in real-time near microtoroids}\label{sec:exp}

In order for the semiclassical model to be applied to our falling atom experiments, we must simulate the atom detection processes. In particular, in~\cite{Alton:2010}, falling Cs atoms are detected with real-time photon counting using a field programmable gate array (FPGA), with subsequent probe modulation triggered by atom detection.

A microtoroidal cavity with frequency $\omega_{\rm c}$ is locked near the $6S_{1/2}, F=4 \rightarrow 6P_{3/2}, F=5$ atomic transition of Cs at $\omega_{\rm a}^0$ at desired detuning $\Delta_{\rm ca} = \omega_{\rm c} - \omega_{\rm a}^{(0)} $.  Fiber-cavity coupling is tuned to critical coupling where the bare cavity transmission vanishes, $T \lesssim T_{\min} \simeq 0.01$.  For atom detection, a probe field at frequency $\omega_{\rm p} = \omega_{\rm c}$ and flux $P_{\rm in} \sim 15$ cts/$\mu$s is launched in the fiber taper and the transmitted output power $P_{\rm T}$ is monitored by a series of single photon detectors. Photoelectric events in a running time window of length $\Delta t_{\rm th}$ are counted and compared to a threshold count $C_{\rm th}$.  A single atom disturbs the critical coupling balance so that $T/T_{\rm min} > 1$, resulting in a burst of photons which correspond to a possible trigger event.  Extensive details of the experimental procedure are given in~\cite{Alton:2010} and the associated Supplementary Information.

Whereas only a single atom is required to produce a trigger, spectral and temporal data are accumulated over many thousands of trigger events since each individual atom is only coupled to the cavity for a few microseconds.  Simulation is a valuable technique to disentangle atomic dynamics from the aggregate data and offer insights into the atomic motion which underlies the experimental measurements.

\subsection{Simulation procedure}

Central to our simulations is the generation of a set of $N$ representative atomic trajectories for the experimental conditions of atoms falling past a microtoroid fulfilling the criteria for real-time detection.  Since experimental triggering is stochastic, the trajectory set is generated randomly as well.  For each desired collection of experimental parameters $\mathcal{P}$, a set of semiclassical atomic trajectories $\{\vect{r}_j(t)\}_{\mathcal{P}}$ is generated that satisfies the detection trigger criteria. This ensemble is used to extract the cavity output functions $T(t, \Delta_{\rm ap})$ and $R(t, \Delta_{\rm ap})$.  For each individual trajectory, $t=0$ is defined to be the time when the trajectory is experimentally triggered by the FPGA.  For each set $\mathcal{P}$, $N$ is chosen large enough for a sufficient ensemble average to be obtained for the final output functions, which is typically at least 400 unique triggered trajectories.

Within each simulation, the probe field is fixed to a given $\omega_{\rm p}$.  Cavity behavior is determined by the parameters $\omega_{\rm c}$, $h$, $\kappa_{\rm i}$, and $\kappa_{\rm ex}$.  $h$ and $\kappa_{\rm i}$ are determined from measurements of the bare cavity with no atoms present.  Low-bandwidth fluctuations in $\kappa_{\rm ex}$ and $\omega_{\rm c}$ from mechanical vibration and temperature locking are modeled as normally distributed random variations with standard deviations of 3 MHz and 1.5 MHz, respectively, that are fixed once for the duration of each simulated trajectory.  Similar to the experimental procedure, we impose that the bare-cavity output flux is less than 0.4 cts/$\mu$s at critical coupling and on resonance.  This rate would be identically zero for $\Delta_{\rm cp} = 0$ and critical coupling in the absence of these fluctuations.  If the noise threshold is not met, then the particular trajectory is thrown out as it would have been in experiments.

The atomic cloud is characterized by its temperature, size, and its height above the microtoroid. Its shape is assumed to be Gaussian in each direction with parameters determined by florescence imaging.  For each simulation loop, an initial atomic position $\vect{r}_{\rm in}$ is selected from the cloud and the initial velocity $\vect{v}_{\rm in}$ is selected from a Maxwell-Boltzmann distribution of temperature $T$.  The trajectory is propagated forward in time under the influence of gravity until it crosses the toroid equatorial plane at $z=0$.  Only trajectories which pass within $1$ $\mu$m of the toroid surface at $z=0$ are kept as a candidate for detection, as other atoms couple too weakly for triggering. Once an acceptable set of initial conditions is obtained, the trajectory $\vect{r}(t)$ is calculated over a 50 $\mu$s time window around its crossing of $z=0$, this time with the gravity, optical dipole forces, and surface interactions included.  As the atom moves through the cavity mode, the atom-cavity coupling $g$, level shifts, decay rates, and forces change with position $\vect{r}$, causing deviations of the trajectory from the preliminary free-fall trajectory. If the atom crashes into the surface of the toroid, then the coupling is set to $g=0$ onwards and the trajectory effectively ends (except for random `noise' photon counts arising from the non-zero background transmission).

Using $\vect{r}(t)$ and the steady-state expressions for the fields (section~\ref{Sec:theory}), we find the transmission $T(t)$.  The photon count record $C_i(t_j)$ on each photodetector $i$ for time step $t_j$ is generated from a time-dependent Poisson process with mean count per bin of $\overline{C_i}(t_j) = T(t_j) P_{\rm in}\Delta t$, where $\Delta t = t_{j+1}-t_j = 1$ ns and $P_{\rm in}$ is the input flux.  Since the typical flux is $P_{\rm in} \sim 10$ MHz and the timescale of quantum correlations is $\sim10$ ns, the photon count process is assumed to be Poissonian on the relevant timescale of a few hundred nanoseconds for atom detection.  The count record $C_i(t_j)$ is compared to the desired threshold of $C_{\rm th}$ in a time window $\Delta t_{\rm th}$~\cite{Alton:2010}.  If the trigger condition is met, the initial conditions $\vect{r}_{{\rm in, }j }$, $\vect{v}_{{\rm in, }j }$, the random cavity parameters $\omega_{\rm c}$ and $\kappa_{\rm ex}$, and the random number seed used to generate $\vect{W}^i$ for diffusion processes are stored for later use.  The semiclassical trajectory $\vect{r}_j(t)$ can be fully reconstructed from these parameters.  The time coordinate is shifted so that the trigger event occurs at $t=0$.  This process is repeated to acquire $N$ triggered trajectories.

Cavity output functions such as the experimentally measurable transmission $T_{\rm exp}(t,\mathcal{P})$ for each simulation parameter set $\mathcal{P}$ are calculated from the set of trajectories $\{\vect{r}_j(t)\}$:
\begin{equation}\label{eq:T}
T_{\rm exp}(t,\mathcal{P}) = \frac{1}{N} \sum_j^N T(\vect{r}_j(t),\mathcal{P})
\end{equation}
Reflection coefficients $R_{\rm exp}(t,\mathcal{P})$ are calculated similarly.  Spectra are calculated by averaging output powers over a time window $t_1 < t < t_2$ for each probe frequency $\omega_{\rm p}$.  The times $t_1$ and $t_2$ are chosen to be the same as in our experiments, which is typically $t_1 = 250$ ns and $t_2$ = 750 ns.  The set of triggered trajectories $\{\vect{r}_j(t)\}$ is valid for a given set of conditions $\mathcal{P}$ and detection criteria $\{C_{\rm th}, \Delta t_{\rm th}\}$ until the trigger at $t=0$.  In experiments, the probe frequency $\omega_{\rm p}$ can be changed in power and detuning upon FPGA trigger. Although the same set of trajectories is valid before $t=0$ for each detuning, the trajectory set must be recalculated for $t >0$ for each probe detuning to mimic experimental conditions for spectral measurements.  A numerical solution of the master equation in a number state basis is used for calculation of $T(t)$ in~\eref{eq:T}; the linearized model is only used to calculate the trajectory $\vect{r}(t)$ and efficiently generate triggers.

Whereas experiments give access only to ensemble averaged output functions, simulations contain the full trajectory paths. Provided that the simulation offers a reasonable approximation of the true ensemble of trajectories, then these results provide a window into the atomic dynamics underlying the cQED measurement of falling atoms which are not readily clear from observations.

\subsection{Simulation distributions}

\begin{figure}[tbh]
  \begin{center}
    \includegraphics{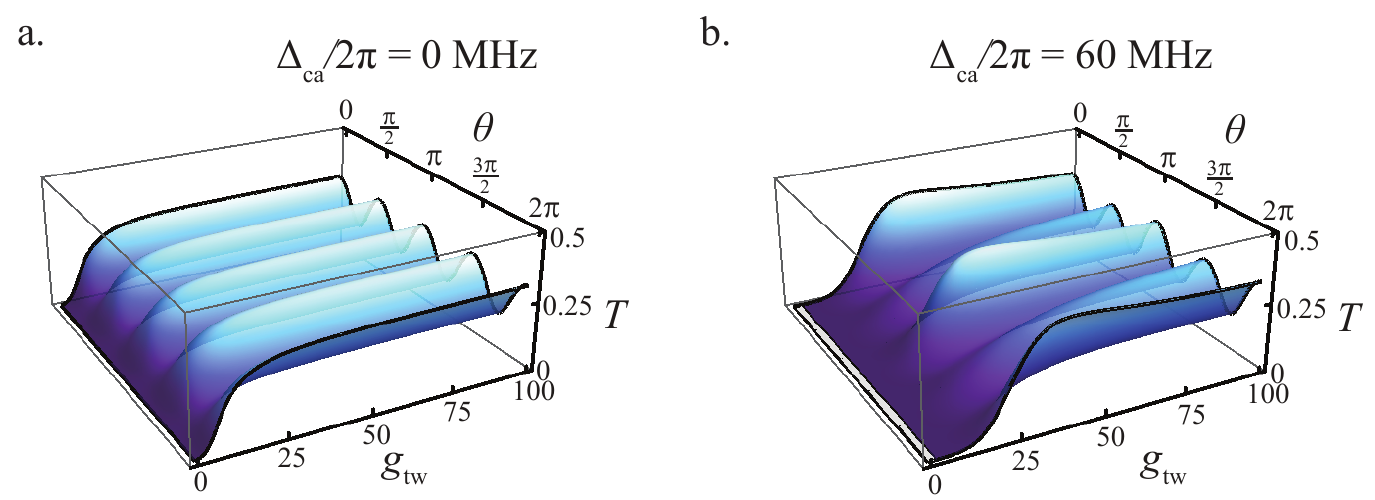}
  \end{center} \vspace{-0.5cm}
\caption{Plots of $T(g_{\rm tw}, \theta, \mathcal{P})$ for (a) and (b) calculated numerically from~\eref{eq:MasterEquation}.  Atoms with higher $g_{\rm tw}$ generally have higher $T$ and a larger probability for detection.  The variation of $T$ with $\theta$ is evident, with a different periodicity for the two cavity detunings.  } \label{fig:Tfunc}
\end{figure}

The experimentally measurable cavity transmission $T_{\rm exp}(t)$ is obtained in~\eref{eq:T} as an average over the trajectory set $\{\vect{r}_j(t)\}$ at each time $t$.  Eq.~\eref{eq:T} can formally be written as an integration over the probability distribution of coupling constants at time $t$, $p_{t}(g, \theta)$, for the given experimental parameters $\mathcal{P}$:
\begin{equation}\label{eq:Tint}
T_{\rm exp}(t,\mathcal{P}) = \int \, d g \, d \theta \, T(g, \theta, \mathcal{P}) p_{t}(g, \theta)
\end{equation}
The function $T(g, \theta, \mathcal{P})$ is shown in Fig.~\ref{fig:Tfunc} for the parameters $\mathcal{P}$ relevant to experiments, specifically with $\Delta_{\rm ca}/2\pi = 0, \, 60$ MHz.  For this discussion, we assume all frequencies are fixed and neglect surface shifts.  In this perspective, $T_{\rm exp}(t)$ is not directly related to the trajectory set $\{\vect{r}_j(t)\}$ but rather the probability distribution $p_{t}(g, \theta)$ at time $t$.  The time dependence of $p_{t}$ evolves based on the underlying trajectory ensemble.

\begin{figure}[tbh]
  \begin{center}
\vfill
    \includegraphics{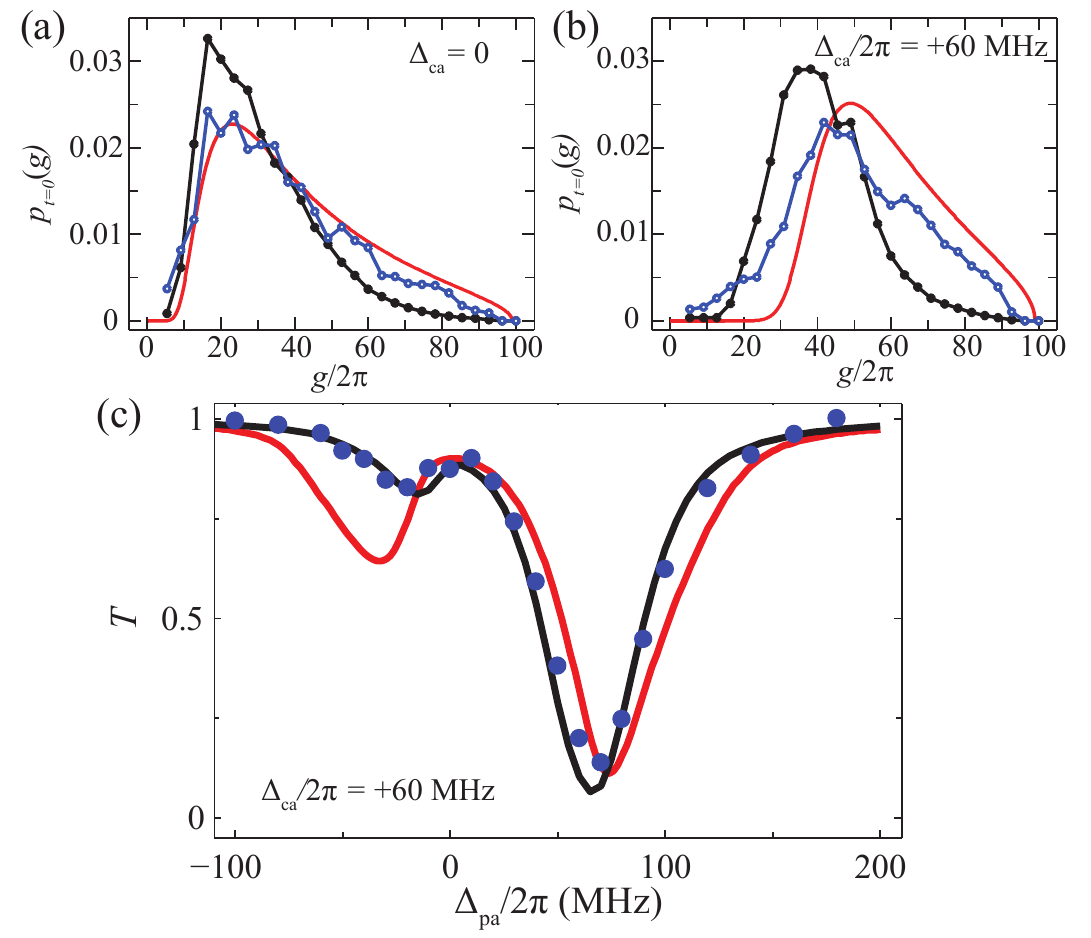}
  \end{center} \vspace{-0.5cm}
\caption{Distributions $p_{t=0}(g)$ of coupling constants calculated for (a) $\Delta_{\rm ca}/2\pi =0$ and (b) $\Delta_{\rm ca}/2\pi = +60$ MHz.  Distributions from the analytic model (red), semiclassical trajectory simulation with no dipole or surface forces (blue), and the simulation with all forces (black) are shown for comparison.  (c) Experimental cQED spectra data for cavity detuning $\Delta_{\rm ca}/2\pi = 60$ MHz (blue points) from~\cite{Alton:2010} plotted with model spectra calculated from the distributions $p_{t=0}(g)$ in panel (b). The red is the analytic model of ~\ref{app:toy} and black is the semiclassical simulation.} \label{fig:dist}
\end{figure}

It is instructive to consider the probability distribution $p_{t}(g, \theta)$ in more detail since it is the formal output of the simulations.  We consider only the distribution $p_{t=0}(g)$ over the coupling parameter $g$ at the trigger time $t=0$ by integrating out the angular dependence.  Through a reasonably simple analytic model (detailed in~\ref{app:toy}), we calculate $p_{t=0}(g)$ and compare to the results of the semiclassical simulation, which includes dipole and surface forces (Fig.~\ref{fig:dist}).  For a cavity on resonance with the atom transition, $\Delta_{\rm ca}/2\pi =0$, the analytic model agrees well with a simulation when dipole and surface forces are not included.   In this case, atom trajectories are nearly straight and vertical near the toroid, and the approximations of~\ref{app:toy} are sufficient.  When the full forces are included in the semiclassical model, the additional forces shift the distribution toward lower $g$.  This effect is more significant for $\Delta_{\rm ca}/2\pi=60$ MHz.  The corresponding experimental cQED spectra confirm that the semiclassical model with dipole and surface forces is necessary to reproduce spectral features in the real-time experiments (Fig.~\ref{fig:dist}c).

\begin{figure}[tbh]
  \begin{center}
    \includegraphics{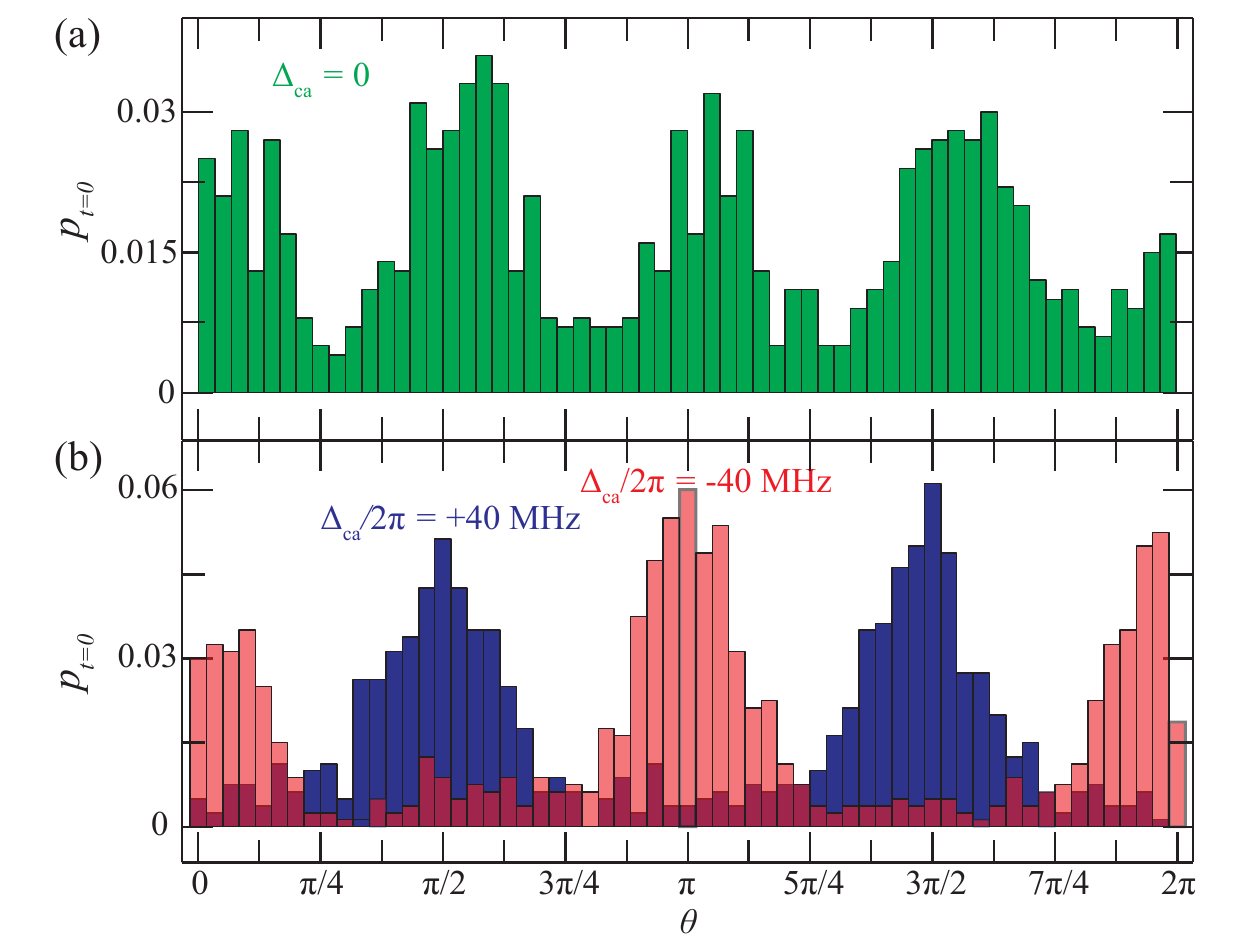}
  \end{center} \vspace{-0.5cm}
\caption{Probability distribution $p_{t=0}(\theta)$ of atomic azimuthal angle $\theta = m \phi \, \textrm{mod} \,2\pi$ at transit detection time $t=0$ presented as histograms of simulation runs.  Shown are the cases for cavity detunings (a) $\Delta_{\rm ca} = 0$ and (b) $\Delta_{\rm ca}/2\pi= \pm 40$ MHz. Normalization is such that the sum across all $\theta$ is unity.} \label{fig:PhiDistribution}
\end{figure}

The cavity transmission $T$ varies as a function of the atomic azimuthal coordinate $\theta = m \phi$, as evident in Fig.~\ref{fig:Tfunc}.  This biases atomic detection towards specific locations around the toroid and leads to a non-uniform angular distribution $p_{t=0}(\theta)$ for atom location at detection.   Fig.~\ref{fig:PhiDistribution} shows distributions of the atomic angular coordinate at the detection trigger $t=0$ for three simulation conditions relevant to the experiments of~\cite{Alton:2010}.  Although averaged spectra do not explicitly measure the coordinate $\theta$, these simulation makes clear that trajectories passing through certain regions around the toroid are preferentially detected.  The phase of the cavity output field depends on $\theta$, suggesting the possibility for future experiments to measure the distribution of Fig.~\ref{fig:PhiDistribution}.

\subsection{Simulated trajectories}

We now turn to the simulated trajectories $\{\vect{r}_j(t)\}$.  In contrast to experiments, in simulations we have the capability of turning certain forces selectively on and off.  In particular, we can adjust the surface potential $U_{\rm s}$ and the dipole forces, referred to symbolically as $U_{\rm d}$ (despite them not being strictly derivable from a potential).  To investigate the effects these optical phenomena have on atomic trajectories, we run simulations for four cases: the full semiclassical model, the model without surface forces ($U_{\rm s} = 0$), the model without dipole forces ($U_{\rm d}=0$), and the model without any radiative forces ($U_{\rm d} = U_{\rm s} = 0$).

\begin{landscape}
\begin{figure}[tbh]
  \begin{center}
    \includegraphics{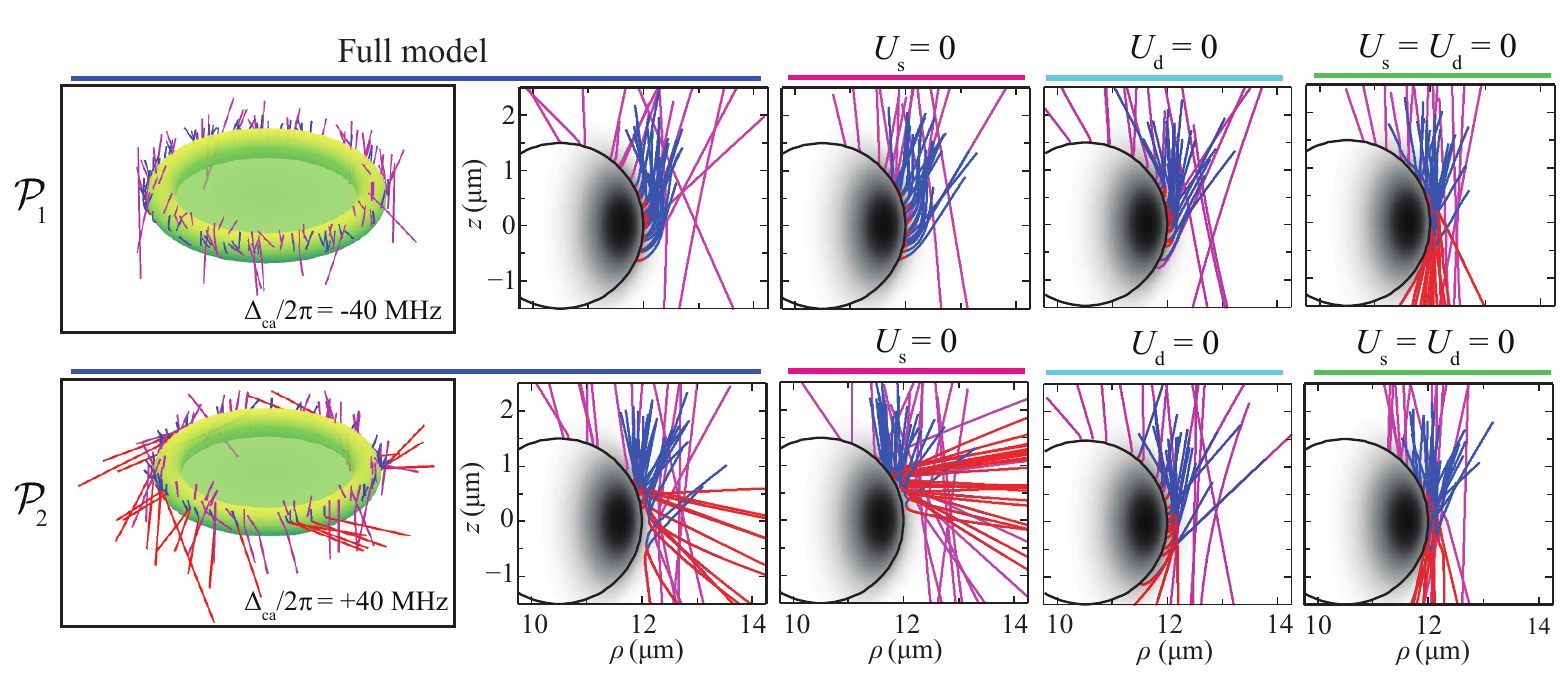}
  \end{center}\vspace{-0.5cm}
\caption{Simulated trajectories for model parameters $\mathcal{P}_{1,2}$ ($\Delta_{\rm ca}/2\pi = \rm 40$ MHz)  plotted for four models of radiative forces: the full semiclassical model, $U_{\rm s}=0$, $U_{\rm d} =0 $, and $U_{\rm s} = U_{\rm d} = 0$. For the full model, a three-dimensional representation is shown, while trajectories are projected onto the two-dimensional $\rho-z$ plane for all conditions.  Magenta trajectories represent \emph{un-triggered} atoms, blue paths are detected atoms for $t<0$ and red paths represent atom trajectories after the trigger for $t>0$.} \label{fig:Trajectories}
\end{figure}
\end{landscape}

Considering conditions relevant to~\cite{Alton:2010}, we plot simulations for two sets of experimental parameters $\mathcal{P}_{1,2}$ in Fig.~\ref{fig:Trajectories}.  For $\mathcal{P}_1$, the cavity is detuned to the red, whereas the cavity is blue-detuned in $\mathcal{P}_2$ ($\Delta_{\rm ca}/2\pi = -40$ MHz for $\mathcal{P}_1$ and $+40$ MHz for $\mathcal{P}_2$).  In each set of conditions, the probe field is on resonance with the cavity for high signal-to-noise atom detection ($\Delta_{\rm cp} =0$) and the average bare-cavity mode population of $a$ is $\approx 0.05$ photons.  The toroid cavity parameters are those of~\cite{Alton:2010}, $\{g_{\rm max}, h, \kappa_{\rm in}, \kappa_{\rm ex}\}/2\pi = \{100, 11, 13, 17\}$ MHz.  Comparing the full model, we see that trajectories for $\mathcal{P}_1$ primarily crash into the surface,  whereas those from $\mathcal{P}_2$ both crash and are repelled from the toroid. This asymmetry is due to the repulsive or attractive dipole force for different probe detunings relative to the atomic transition.  The largest effect of turning surface forces \emph{off} is seen in the blue-detuned trajectories, which have a lower crash rate when $U_{\rm s}=0$.  With $U_{\rm d}=0$, both $\mathcal{P}_{1}$ and $\mathcal{P}_{2}$ trajectories look nominally the same; the detuning $\Delta_{ca}$ only affects cQED spectra, with a minor imperceptible effect arising from CP potentials initially shifting the atomic transition either closer to (red) or further from (blue) the cavity field.

In addition to the qualitative differences in detected atom trajectories summarized here, the effects of $U_{\rm d}$ and $U_{\rm s}$ are evident in the experimental quantities $T_{\rm exp}(t)$ and $R_{\rm exp}(t)$.  Since here we focus specifically on trajectory calculations, the reader is referred to~\cite{Alton:2010} for detailed comparisons of spectral and temporal simulations to experimental data.

\section{Trapping atoms in the evanescent field of a microtoroid}\label{sec:trap}

Our trajectory simulation can be extended to study trapping of atoms in a two-color evanescent far off-resonant trap (eFORT) near a microtoroidal resonator.  An evanescent field trap takes advantage of the wavelength dependence of scale lengths for the optical dipole force of two optical fields with frequencies far-detuned from the atomic transition to limit scattering ~\cite{Cook:1982, Ovchinnikov:1991, Dowling:1996}.  The relative powers of the two fields are set so that near the surface, the blue-detuned, repulsive field is stronger than a red-detuned attractive field.  As each field falls off with a decay constant of roughly $\lambdabar = 2 \pi/\lambda$, at some distance, the red, attractive field will dominate and the atom will be attracted to the surface forming a potential minimum.  Recently, evanescent fields have been harnessed to trap atoms in a two-color eFORT around a tapered optical fiber~\cite{Vetsch:2010}, where the fiber enables efficient optical access to deliver both high intensity trapping fields and weaker probe fields to the trapped atoms in a single structure.  The tapered fiber can be positioned as desired, bringing the trapped atoms near a device for atomic coupling.
	
The tapered nanofiber eFORT is a remarkable achievement toward integrating atom traps with solid-state resonators, but the nanofiber scheme does not allow direct integration with a cavity for achieving strong, coherent coupling between light and trapped atoms.    Another disadvantage is that trap depth is limited by the large total power required to achieve trapping with evanescent fields.  The high quality factors and monolithic structure of WGM resonators allow evanescent field traps free from these problems while maintaining efficient optical access from tapered fiber coupling.  Two-color evanescent field traps in WGM resonators have been analyzed in detail for spheres~\cite{Vernooy:1997} and microdisks~\cite{Rosenblit:2006}.  In this section, we extend our simulations of atoms in the evanescent field of a microtoroid to an eFORT that can capture single falling Cs atoms triggered upon an atom detection event.

Unlike nanofibers, a microtoroid cannot be placed directly in a magneto-optical trap for a source of cold atoms.  As shown in~\cite{Alton:2010}, we have the experimental capability to detect a single atom falling by a microtoroid and trigger optical fields while that atom remains coupled to the cavity mode.  The semiclassical simulations described here are ideal for investigating the capture of falling atoms in a trap triggered upon experimental atom detection.

\begin{figure}[tb]
  \begin{center}
    \includegraphics{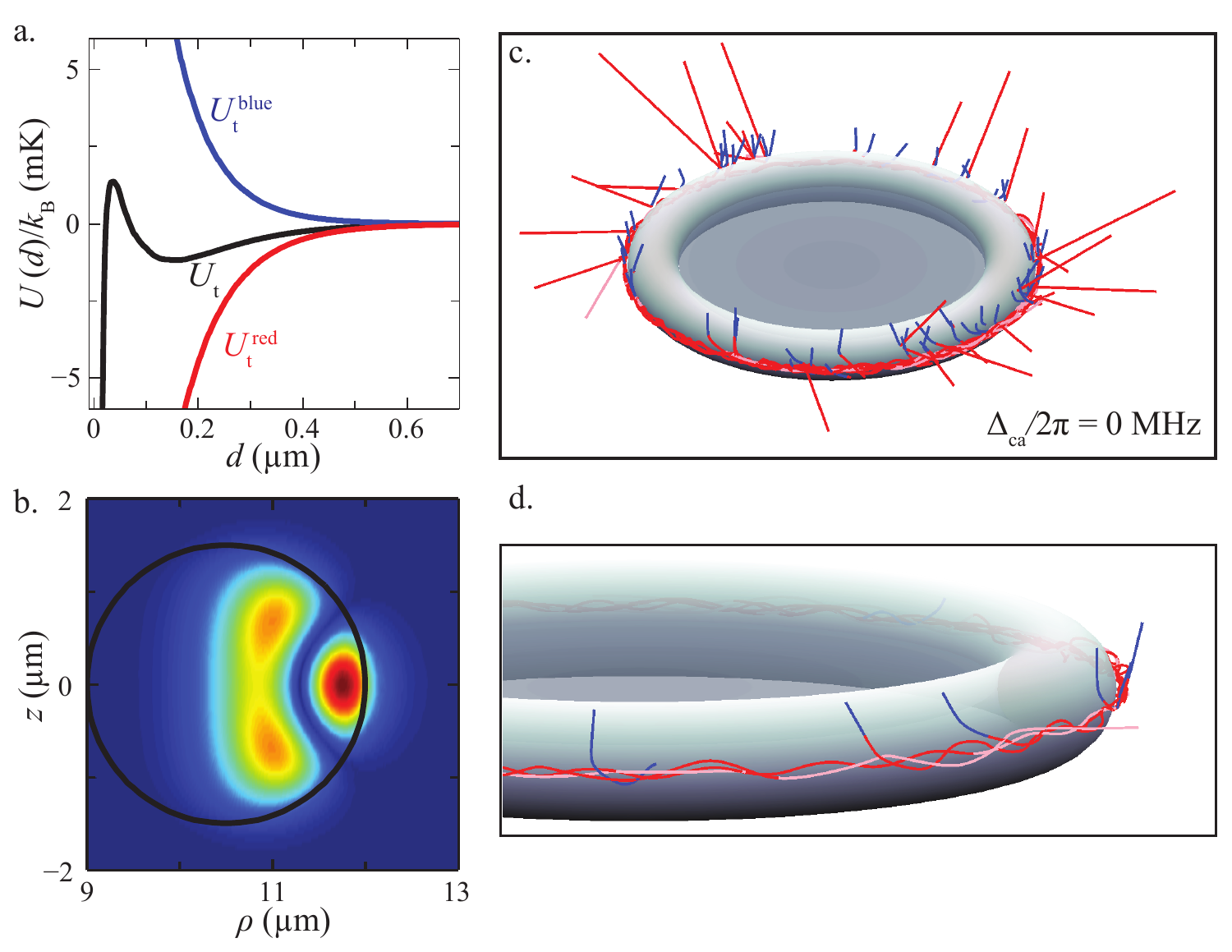}
  \end{center}\vspace{-0.5cm}
\caption{(a) The trapping potential $U_{\rm t}$ along the $z=0$ axis with the CP potential included.  Also shown are the red and blue evanescent potentials of the two trapping modes, $U_{\rm t}$, respectively.  (b) The mode function used in $U_{\rm t}$ for the 898 nm mode with $m=106$. (c) Simulated trajectories for trapping simulations with a eFORT $U_{\rm t}$ triggered ``on" by atom detection at $t = 0$ with $\Delta_{\rm ca} =0$.  Falling atoms with the FORT beams ``off" $(t < 0)$ are colored blue, whereas trajectories after the trap is triggered are red.  Trajectories are colored pink for  $t > 50$ $\mu$s to illustrate the timescale.  Roughly $25\%$ of the triggered trajectories become trapped.   (d) Same as (c) showing only the trapped trajectories and a clearer view of atom orbits in the evanescent trap.} \label{fig:trap}
\end{figure}

We add an additional eFORT potential $U_{\rm t}$ to our semiclassical trajectory model in addition to the dipole forces and surface potential $U_{\rm s}$.  For our simulation, $U_{\rm t}$ is formed from a red (blue)-detuned mode near 898 nm (848 nm) with powers $\sim 50$ $\mu$W to give a trap depth of $\sim 1.5$ mK at $d \sim 150$ nm from the surface (Fig.~\ref{fig:trap}a).   The red (blue) fields interact primarily with the $6S_{1/2} \rightarrow 6P_{1/2}$ $(6S_{1/2} \rightarrow 6P_{3/2})$ transition.   The trap depth is limited by the total power in vacuum that can propagate in the tapered fibers of~\cite{Alton:2010}.  Power handling can be improved with specific attention to taper cleanliness, so with experimental care the trap depth can be increased reasonably from the discussion here, although we simulate under the conditions given to illustrate that this trap is already experimentally accessible.

The difference in vertical scale lengths ($\psi_0$ in~\eref{eq:approxf}) for modes of different wavelength leads to a trap that is not fully confined if both the red- and blue-detuned trap modes are of the lowest order (as in Fig.~\ref{subsec:modes}b).  As $|\psi|$ increases, the repulsive blue-detuned light weakens faster than the red-detuned field, and atoms can crash into the toroid surface.  This problem is alleviated by exciting a higher-order mode for the 898 nm light, as shown in Fig.~\ref{fig:trap}b.  The modal pattern confines atoms near $z=0$ and prevents trap leakage along $\psi$.  This problem is not present in the microdisk eFORT of~\cite{Rosenblit:2006} because the optical mode extent is determined by structural confinement and not the optical scale length.  Use of a higher-order mode was also used to form an atom-gallery in a microsphere~\cite{Vernooy:1997}.

During the detection phase of the simulation, $U_{\rm t}=0$.  At $t=0$ conditioned on an atom detection trigger, $U_{\rm t}$ is turned on.  The kinetic energy of an atom with typical fall velocity of $v\sim0.2$ m/s is equivalent to 0.3 mK, so a 1.5 mK trap is sufficiently deep to capture an atom if it is triggered near the trap potential minimum.  Defining a trapped trajectory to be one such that the atom has $g/2\pi > 5$ MHz at $t = 10$ $\mu$s, approximately $25\%$ of triggered atom trajectories are captured when the trapping potential is turned on.  Simulated trapping times exceed 50 $\mu$s, limited not by heating from trapping light but by the radiation pressure from the unbalanced traveling whispering-gallery modes of nearly-resonant optical probe field.  This probe field can be turned off so that the atoms remain trapped beyond the simulation time.

In contrast to the standing-wave structure of a typical eFORT or Fabry-Perot cavity trap~\cite{Ye:1999}, microtoroidal resonators offer the tantalizing possibility of radially confining an atom in a circular orbit around the toroid~\cite{ Vernooy:1997, Mabuchi:1994}.  The $U_{\rm t}=0$ outlined here does not confine the atoms azimuthally, forming circular atom-gallery orbits around the microtoroid~\cite{Mabuchi:1994} (Fig.~\ref{fig:trap}c,d).  In the same manner as~\cite{Vetsch:2010}, a localized trap can be achieved by exciting a red-detuned standing wave for three-dimensional trap confinement.  

This trapping simulation outlines how real-time atom detection can be utilized to trap a falling atom in a microtoroidal eFORT.  In practice, microtoroidal traps present some serious practical challenges.  Notably, because the trap quality is sensitive to the particular whispering-gallery mode, the excited optical mode must be experimentally controlled.  The success of an eFORT for Cs atoms around a tapered nanofiber~\cite{Vetsch:2010} strongly suggests that similar trap performance might be achieved for an eFORT around a high-\textit{Q} WGM cavity, localizing atoms in a region of strong coupling to a microresonator.

\section{Conclusion}

We have presented simulations of atomic motion near a dielectric surface in the regime of strong coupling to a cavity with weak atomic excitation.  As required by experimental distance scales, this simulation includes surface interactions, which manifest through transition level shifts and center-of-mass Casimir-Polder forces.  Analysis of the simulated trajectories gives insight into the atomic motion underlying experimental measurements of ensemble-averaged spectral and temporal measurements for single atoms detected in real-time.  We have adapted our simulations to investigate the capturing of atoms in an evanescent field far off-resonant optical trap in a microtoroid.  Our simulations suggest that falling atoms can be captured into an eFORT around a microtoroid, offering an experimental route towards trapping a single atom in atom-chip trap in a regime with both strong cQED interactions and significant Casimir-Polder forces simultaneously.  In this system, the sensitivity afforded by coherent cQED can be used not only for atom-chip devices, but also as a tool for precision measurements of optical phenomena near surfaces.

\ack
We acknowledge support from NSF, DoD NSSEFF program, Northrop Grumman Aerospace Systems, and previous funding from ARO and IARPA.  N.P.S. acknowledges support of the Caltech Tolman Postdoctoral Fellowship.

\appendix

\section{Calculating the Polarizability and Dielectric Response Functions}\label{app:response}

Evaluation of Casimir-Polder interactions of atoms with the surface of the dielectric resonator requires evaluation of the atomic polarizability and of the dielectric function as functions of a complex frequency.  Here we outline our analytic model of the complex dielectric function for SiO$_{2}$ and the atomic polarizability of Cesium atoms in the ground and excited states.

The complex dielectric function $\epsilon(\omega) = \epsilon_1 + i \epsilon_2$ is modeled using a Lorentz oscillator model of the real and imaginary parts of the response function to analytically introduce frequency dependence and enforce causality,
\begin{equation}\label{Eq:LorentzOsc}
\epsilon(\omega) = \epsilon_\infty + \sum_j \frac{f_j}{(\omega_j^2 - \omega^2)^2 +\omega^2 \gamma_j^2}\left( (\omega_j^2 - \omega^2) + i\omega \gamma_j\right)
\end{equation}
Here, $\omega_j$ is the resonance frequency, $\gamma_j$ is the damping coefficient, and $f_j$ is the oscillator strength for each oscillator in the model. $\epsilon_\infty = \epsilon(\omega \rightarrow \infty) = 1$. $\epsilon$ can be expressed in terms of the complex index of refraction $\tilde{n} = n + i\kappa$ as $\epsilon = \tilde{n}^2 = n^2-\kappa^2 + 2 i n \kappa$, where $n$ is the refractive index and $\kappa$ is the extinction coefficient.  Experimental data for $\tilde{n}$ for SiO$_{2}$ is available over a wide frequency range~\cite{Philipp:1985}, which is used to fit the parameters of ~\eref{Eq:LorentzOsc} for a seven-oscillator model ($j = 1-7$).  Using the analytic form of~\eref{Eq:LorentzOsc}, the dielectric function can readily be evaluated over complex frequencies as shown in Fig.~\ref{Fig-Response}.

The frequency-dependent atomic polarizability $\alpha_s(\omega)$ for Cesium in a  state  $s$ is calculated as a sum over transitions of the form,
\begin{equation}\label{Eq:alpha}
\alpha_s(\omega) = \sum_n \frac{e^2 f_{ns}}{m_e} \frac{1}{\omega_{ns}^2 - \omega^2},
\end{equation}
where $e$ is the electron charge, $m_e$ is the electron mass, $\omega_{ns}$ is the transition frequency, and $f_{ns}$ is the signed oscillator strength for the transition of state $n$ to the state $s$ ($f_{ns} > 0$ if state $n$ is above $s$ in energy).  A more complete expression for the response function $\alpha(\omega)$ should include damping coefficients given by the transition linewidths. Since our calculations involve integrals over infinite frequency on the imaginary axis and atomic linewidths are generally narrow with respect to transition frequencies, we assume that the off-resonant form given by~\eref{Eq:alpha} without damping is sufficient.  We also note that this expression does not account for the differences between magnetic sublevels and hyperfine splitting, which again represent small corrections when these expressions are integrated over the imaginary frequency axis.  The general form of~\eref{Eq:alpha} applies to the polarizabilities for both the $6S_{1/2}$ ground state and the $6P_{3/2}$ excited state, with an additional tensor polarizability for the $6P_{3/2}$ state.

The total atomic polarizability is composed of contributions from valence electron transitions ($\alpha_{\rm v}$) and high-energy electron transitions from the core shells to the continuum ($\alpha_{\rm c}$), such that $\alpha =  \alpha_{\rm v} + \alpha_{\rm c}$.  The valence polarizability $\alpha_{\rm v}$ constitutes 96\% of the total static polarizability~\cite{Derevianko:1999} in Cs, with $\alpha_{\rm c}$ only significant at high frequencies.  We take $\alpha_{\rm c}$ to be the same for both the ground and excited states of Cs, whereas $\alpha_{v}$ is obviously sensitive to the different electronic transition manifolds for $6S_{1/2}$ and $6P_{3/2}$ states. Valence electron oscillator strengths and transition frequencies are tabulated in many sources~\cite{Norcross:1973, Laplanche:1983}. Our estimate of $\alpha_{\rm v}(\omega)$ for the ground state includes all $6S_{1/2}\rightarrow NP_{1/2}$ and $6S_{1/2}\rightarrow NP_{3/2}$ transitions, with $N=6 - 11$.  For the excited state, $\alpha_{\rm v}(\omega)$ is calculated using $6P_{3/2} \rightarrow (6-15)S_{1/2}$, $6P_{3/2} \rightarrow (5-11)D_{3/2}$, and $6P_{3/2} \rightarrow (5-11)D_{5/2}$ transitions.  Tensor polarizability contributions sum to zero when averaged over all angular momentum sublevels~\cite{Kien:2005}.  In agreement with~\cite{Derevianko:1999}, our calculation of $\alpha_{\rm v}$ comprises about 95\% of the total static polarizability.

For simplicity, all core electron transitions are lumped into a single high-frequency term of the form used in~\eref{Eq:alpha}. This term contains two free parameters, $f_{\rm core}$ and $\omega_{\rm core}$, which are found from the following two conditions.  Using the calculation of $\alpha_{\rm v}(\omega)$ for the Cs ground state, we enforce that the ground state static polarizability $\alpha(\omega\rightarrow 0)$ matches the known value calculated theoretically~\cite{Amini:2003} $\alpha(0) = 5.942\times10^{-23}$ cm$^3$.  We also ensure that the ground state LJ constant for a Cs atom near a metallic surface agrees with the known value~\cite{Derevianko:1999, Johnson:2004} $C_3 = -\frac{\hbar}{4\pi d^3}\int_0^{\infty} \alpha(i\xi)d\xi = 4.4 \cdot h$ kHz $\mu$m$^3$. These conditions are sufficient to fix the two free parameters in $\alpha_{\rm c}(\omega)$ for this single oscillator core model, although the high-frequency structure of the core polarizability is lost.  For the excited state calculation, we use the same $\alpha_{\rm c}(\omega)$.

\section{Analytic model of falling atom detection distributions}\label{app:toy}

Here we develop an analytic model of the distribution $p_{\rm fall}(g, \theta)$ of coupling parameters $g$ and azimuthal coordinate $\theta= m \phi $.  Atoms are assumed to fall at constant vertical velocity with no forces, in contrast to the more complete semiclassical trajectories used in this manuscript to generate $p_{t}(g)$.  An abbreviated description of this model appears in the Supplementary Information of \cite{Alton:2010}.

The linearized steady-state cavity transmission $T(\Delta_{\rm ap}, g(\vect{r}))$ is a known function of $\Delta_{\rm ap}$ and $\vect{r}$.  We only consider the lowest order mode where the cavity mode function is approximately Gaussian in $z$ and exponential in distance from the surface $d$.  The approximate temporal behavior of the coupling constant $g$ for a single trajectory is,
\begin{equation}\label{eq:gtrajectory}
g(\rho, z(t)) = g_{\rm c}(\rho) e^{-\left(z(t)/z_0\right)^2} .
\end{equation}
where $g_{\rm c}(\rho)$ is the maximum value of the $g$ at the closest  approach of its trajectory ($z=0$), $z_0$ is a characteristic width assumed to be independent of $\rho$, and $z(t) = - v t$.  $g_{\rm c}(\rho)$ decays exponentially from the maximum $g_{\rm max}$ at the toroid surface, $g_{\rm c}(\rho) \sim g_{\rm max} e^{-(\rho-D_{\rm p})/\lambdabar_0}$.  The transmission $T$ and hence the detection probability depend on $\theta$; in general, if atoms fall uniformly around the toroid, the most numerous trajectories detected will be at the values of $\theta$ which maximize $T(\theta)$ for the cavity parameters of interest ($\theta = \pi/2$ for $\Delta_{\rm ca}/2\pi = +40$ MHz, for example, as in Fig.~\ref{fig:PhiDistribution}).

The probability density function for the full ensemble of detected falling atoms $p_{\rm fall}(g, \theta)$ can be estimated as the product of the probability of any atom having a particular $g$ and the probability of a trigger event occurring for an atom with coupling $g$,
\begin{equation}\label{eq:fallprob}
p_{\rm fall}(g, \theta) \sim p_{\rm atom}(g)p_{\rm trigger}(g, \theta).
\end{equation}
An atom transit is triggered when the total detected photon counts exceeds a threshold number, $C_{\rm th}$, within a detection time window $\Delta t_{\rm th}$.  For a probe beam of input flux $P_{\rm in}$, the mean counts in this window are $\overline{C} = T(g, \theta) P_{\rm in} \Delta t_{\rm th}$.  This expression assumes that the atom is moving slowly so that the $T(g, \theta)$ at trigger event is the only $T(g, \theta)$ that contributes to the detection probability.   The detection probability $p_{{\rm trigger}}(g, \theta)$ is estimated from a Poisson distribution of mean count $\overline{C}$.

From~\eref{eq:gtrajectory}, $p_{\rm atom}(g)$ can be written as a product of the probability $p(g|g_{\rm c})$ of an atom in a trajectory with a given  $g_{\rm c}$ to have coupling $g$ and the probability of a trajectory to have that $g_{\rm c}$, $p_{\rm max}(g_{\rm c})$, integrated over all $g_{\rm c}$,
\begin{equation}\label{Eq:Patom}	
p_{\rm atom}(g) = \int_{g}^{g_{\rm max}} p(g|g_{\rm c}) p_{\rm max}(g_{\rm c}) \,d g_{\rm c}
\end{equation}
The integral has limits from $g$ to $g_{\rm max}$ since $g_{\rm c}$ cannot be smaller than $g$.

For atoms falling uniformly over the $\rho-\phi$ plane, $p_{\rm max}(g_{\rm c}) \,d g_{\rm c}$ is proportional to the area of a ring of radius $\rho$ and thickness $d\rho$, $ p_{\rm max}(g_{\rm c}) \,d g_{\rm c} \sim 2 \pi \rho\, d \rho$.  Using $g_{\rm c}(\rho) \sim e^{-\left(\rho - D_{\rm p}/2\right)/\lambdabar_0}$, $\frac{d g_{\rm c}}{g_{\rm c}} \sim -\frac{d \rho}{\lambdabar_0}$. Hence, $p_{\textrm{max}}(g_{\rm c}) \sim 1/g_{\rm c}$ for $(\rho-D_{\rm p}/2) \ll D_{\rm p}/2$.  To find $p(g|g_{\rm c})$ we note that that the probability is proportional to the time an atom in the trajectory is at a particular $g$.   From~\eref{eq:gtrajectory} for a constant velocity $v$, this trajectory is Gaussian and the relative probability must be proportional to $d z$.  Finding the differential as a function of $g$ gives $p(g|g_{\rm c}) \propto d z \sim \frac{1}{g \sqrt{\ln(g_{\rm c}/g)}}$.

Putting the results together in~\eref{Eq:Patom} gives
\begin{equation}\label{Eq:Patom2}
p_{\rm atom}(g) \sim \int_{g}^{g_{\rm max}} \frac{1}{g g_{\rm c}} \frac{d g_{\rm c}}{\sqrt{\ln(g_{\rm c}/g)}}  \sim \frac{\sqrt{\ln \left(\frac{{g_{\rm max}}}{g}\right)}}{g}
\end{equation}
This result diverges as $g$ goes to zero since there are infinite transits with small $g_{\rm c}$ and infinite time for atoms with small $g$ for any transit regardless of $g_{\rm c}$ for $t\rightarrow \pm \infty$.  This divergence is not problematic in calculating~\eref{eq:fallprob} since $p_{\textrm{trigger}}(g, \theta)$ cuts off for low $g$ faster 
than the logarithmic divergence in $p_{\rm atom}(g)$.

The spectrum for given experimental parameters as a function of probe detuning $\Delta_{\rm ap} = \omega_{p} - \omega_{\rm a}^{(0)}$ can be written as:
\begin{equation}\label{Eq:spectrum}
T(\Delta_{\rm ap}) = \int_0^{g_{\rm max}} T(\Delta_{\rm ap}, g, \theta) p_{\rm fall}(g, \theta) \,d g \, d \theta
\end{equation}
where the normalization of $p_{\rm fall }(g, \theta)$ is chosen such that
\begin{equation}
\int_{0}^{g_{\rm max}} p_{\rm fall}(g, \theta) \,d g \, d \theta = 1
\end{equation}
The overall probability of $g$, $p_{\rm fall}(g)$ independent of $\theta$, is found by integrating over $\theta$.  In practice, $p_{\rm fall}(g)$ is quite similar to $p_{\rm fall}(g, \theta)$ evaluated for the $\theta$ which maximizes the transmission.  Fig.~\ref{fig:dist} compares this simple model for $p_{\rm fall}(g)$ with the equivalent distribution from the semiclassical trajectory simulation, $p_{t=0}(g)$.

\section*{References}

   \providecommand{\newblock}{}

\end{document}